\documentclass[aps,pra,reprint,twocolumn,superscriptaddress,showpacs,showkeys,superscriptaddress,longbibliography]{revtex4-2}
\usepackage{dcolumn}    
\usepackage{ifpdf}
\usepackage{amssymb,lineno,amsfonts}
\usepackage{graphicx}   
\usepackage{dcolumn}    
\usepackage{bm}         
\usepackage{mathrsfs}
\usepackage{upgreek}
\usepackage{mathtools}
\usepackage{epstopdf}
\usepackage{setspace}
\usepackage{hyperref}
\usepackage{float}
\usepackage{amsmath}
\usepackage{placeins}
\usepackage{bbm}
\usepackage[normalem]{ulem}
\usepackage[utf8]{inputenc}
\usepackage[usenames,dvipsnames]{xcolor}
\usepackage[matrix,frame,arrow]{xypic}
\usepackage[mathscr]{euscript}
\usepackage[]{qcircuit}
\usepackage[normalem]{ulem}

\definecolor{med-blue}{RGB}{25,25,112}
\hypersetup{colorlinks, linkcolor={red},citecolor={blue}, urlcolor={MidnightBlue}}

\newcommand{\ket}[1]{\vert{#1}\rangle}
\newcommand{\bra}[1]{\langle{#1}\vert}

\newcommand{\expv}[1]{\langle{#1}\rangle}

\begin{document}
	\title{NMR investigations of Dynamical Tunneling in Spin Systems}
	\author{V. R. Krithika}
	\email{krithika$_$vr@students.iiserpune.ac.in}
	\affiliation{Department of Physics, 
		Indian Institute of Science Education and Research, Pune 411008, India}
	\affiliation{NMR Research Center, 
		Indian Institute of Science Education and Research, Pune 411008, India}
	\author{M. S. Santhanam}
	\email{santh@iiserpune.ac.in} 
	\affiliation{Department of Physics, 
		Indian Institute of Science Education and Research, Pune 411008, India}
	\author{ T. S. Mahesh}
	\email{mahesh.ts@iiserpune.ac.in}
	\affiliation{Department of Physics, 
		Indian Institute of Science Education and Research, Pune 411008, India}
	\affiliation{NMR Research Center, 
		Indian Institute of Science Education and Research, Pune 411008, India}
	
	\begin{abstract}
        In the usual quantum tunneling, a low-energy quantum particle penetrates across a physical barrier of higher potential energy, by traversing a classically forbidden region, and finally escapes into another region.  In an analogous scenario, a classical particle inside a closed regular region in the phase space is dynamically bound from escaping to other regions of the phase space. Here, the physical potential barrier is replaced by dynamical barriers which separate different regions of the phase space. However, in the quantum regime, the system can overcome such dynamical barriers and escape through them, giving rise to \textit{dynamical tunneling}. In chaotic Hamiltonian systems, dynamical tunneling refers to quantum tunneling between states whose classical limit correspond to symmetry-related regular regions separated by a chaotic zone between which any classical transport is prohibited. Here, an experimental realization of dynamical tunneling in spin systems is reported using nuclear magnetic resonance (NMR) architecture. In particular, dynamical tunneling in quantum kicked tops of spin-1 and spin-3/2 systems using two- and three-qubit NMR registers is investigated. By extracting time-dependent expectation values of the angular momentum operator components, size-dependent tunneling behaviour for various initial states is systematically investigated. Further, by monitoring the adverse effects of dephasing noise on the tunneling oscillations, we assert the importance of quantum coherence in enabling dynamical tunneling.
	\end{abstract}
	
	\maketitle

	\section{Introduction}
	\label{Introduction}
	Quantum tunneling usually refers to the phenomenon by which a wave packet penetrates and transits through a physical potential barrier despite having lesser energy than the barrier height \cite{griffiths2018introduction}. Classically, this is a forbidden process, though it is allowed in a quantum system. The quantum tunneling phenomenon has been studied extensively and has found applications in various fields ranging from nuclear physics, superconductivity, and electronics to microscopy \cite{griffiths2018introduction,Jospehson1962,Shapiro1963,tinkham2004introduction,bedrossian1989demonstration,Mevel1993tunnelionization,ankerhold2007quantum,chen2021introduction}.  
	
	In chaotic Hamiltonian systems, quantum tunneling manifests into a much richer and more complex phenomenon due to the complexity of underlying classical dynamics \cite{tomsovicullmotunneling,keshavamurthy2011dynamical}. 
	Interestingly, it was realized that the quantum tunneling phenomenon can be extended to scenarios even without any physical barrier. In such cases, the potential barriers are replaced by dynamical barriers formed by invariant phase space structures in the classical limit. Hence, this is often called dynamical tunneling and was first studied by Davis and Heller \cite{davis1981quantum,heller1981quantum} in a two-dimensional nonlinear system. Dynamical tunneling happens when a wave packet tunnels between symmetry-related regular regions such as elliptic islands. It is important to note that the regular regions are separated, not necessarily by potential barriers, but by dynamical constraints. A classical particle initialized in one such regular region can never couple with the other, and hence any transport between these regions is forbidden. In a semiclassical sense, these regular regions would contribute to degenerate eigenstates. However, if tunneling is present between these regular classical regions, we expect the disconnected classical regions to be coupled by quantum dynamics and the degeneracy is lifted. This results in characteristic tunneling doublets in the energy spectrum. The corresponding eigenstates are symmetric and anti-symmetric linear combinations of wavefunctions that predominantly localize on these regular regions \cite{tomsovicullmotunneling,davis1981quantum,peres,tomsovic2001tunneling,ankerhold2007quantum}. This can be effectively modelled as a two-state process (a two-level system) involving these nearly-degenerate states.

	It was found that the tunneling rate between the regular regions can be further enhanced if these regions are separated by a sea of chaos \cite{tomsovicullmotunneling}. In this case, the tunneling wave function has an overlap also with the chaotic region, which aids the tunneling process.  In this case dynamical tunneling, termed as chaos-assisted tunneling, can be thought of as a process involving three levels -- the two nearly-degenerate states coupled through an intermediate chaotic state. The chaotic state can be modelled as a typical state drawn from an appropriate random matrix ensemble.  
	It must also be pointed out that a similar mediation by the classical nonlinear resonances, called the resonance assisted tunneling, in near-integrable regime also leads to enhanced tunneling rates between low and high excited states lying within the same nonlinear resonance region \cite{brodierRAT20001,backermushroombilliards,brodier2002resonance,lockregulartochaotic,Eltschka2005,gehlarexperimental,fritzsch}.
	The rate of tunneling in integrable systems comparatively is much slower due to the absence of resonances and chaos. 
	It is evident that quantum tunneling behaviour can be strongly influenced by the underlying classical structures arising from integrability and non-integrability of the systems \cite{tomsovic2001tunneling,keshavamurthy2011dynamical}.
	
	Though dynamical tunneling has  been theoretically explored for the last three decades, experimental demonstrations are far fewer \cite{tomsovicullmotunneling,Dembowski2000,raizen,Steck2001fluctuations,hensinger2001dynamical,Mouchet,Hofferbert2005,Schlagheck2021,VanhaeleNOON2022,UrbashiS2022}. They are limited to essentially two chaotic test-beds, namely, a driven cold atomic cloud \cite{Mouchet,raizen,hensinger2001dynamical} and microwave annular billiards \cite{Dembowski2000,Hofferbert2005}. Despite the popularity of kicked models within the fold of quantum chaos, especially the ones based on spins such as the kicked top model \cite{haake1987classical}, only one experimental demonstration until now has employed kicked systems \cite{chaudhury2009quantum}. A theoretical study of dynamical tunneling in quantum kicked top (QKT) had been reported in Refs. \cite{sanders1989effect} and \cite{Dogra2019}. Ref. \cite{sanders1989effect} showed that in the presence of dynamical tunneling between regular regions, the expectation values of angular momentum operator components display periodic revivals. To our knowledge, this feature has not been explicitly shown through experiments so far.
	
	
	Nuclear Magnetic Resonance (NMR) has been a convenient testbed for quantum simulations and development of methodologies for quantum information processing \cite{cory2000nmr,suter2008spins}.  Previous NMR studies of nonlinear dynamics include investigating bifurcation in a quadrupolar NMR system \cite{araujo2013classical}, realizing QKT with nuclear spin qubits \cite{krithika2019}, phase synchronization in a pair of interacting nuclear spins subjected to an external drive \cite{krithika2022observation}, quantum phase transitions \cite{Suter2005QPT,Fortes2020,Li2020}, out-of-time-order correlations in integrable and non-integrable systems \cite{DuOTOC}, etc.
	In this work, we carry out NMR investigation of dynamical tunneling in a QKT model formulated as a collection of periodically kicked and interacting spins. This model is useful because the approach to classical limit can be attained  by expanding the Hilbert space,  by either increasing the number of spins, or the spin number, or both. Hence, this system provides a convenient route to study dynamical tunneling and pushing it towards the classical limit.  By monitoring the expectation values of the angular momentum operators of the QKT, we performed a systematic experimental investigations into (i) dynamical tunneling in spin systems for different initial states, (ii) system size dependence of tunneling period with two different system sizes, and (iii) effect of dephasing noise on the robustness of tunneling.  
	By observing the prolonged time periods of dynamical tunneling in the larger system, we infer the inverse size-dependence of the phenomenon.  The dephasing noise also resulted in dampening of tunneling amplitudes, which incidentally appears to have relatively stronger effects on the larger system.
	
	The paper is organized as follows. We introduce QKT model and the concept of dynamical tunneling in spin systems in Sec. \ref{Theory}. We explain the methodologies of NMR experiments in Sec. \ref{NMRExp}, followed by results in Sec. \ref{Res}, and finally conclude in Sec. \ref{Conc}.

	\section{Dynamical tunneling in spin systems}
	\label{Theory}
	\subsection{Quantum Kicked Top (QKT) model}
	The QKT model of a spin-$j$ system is described by the Hamiltonian (with $\hbar$ set to unity) \cite{haake1987classical,sanders1989effect}
    \begin{equation}
        H_{\mathrm{qkt}} = \frac{\pi}{2}J_y\sum_n\delta(t-n\tau) + \frac{k}{2j}J_z^2,
    \end{equation}
    where $J_\alpha$ with $\alpha=x,y,z$ are components of the angular momentum operator. The first term describes an instantaneous kick about the $y$-axis which brings about a rotation of $\pi/2$ angle, and the second term characterized by the chaoticity parameter $k$ describes a nonlinear torsion about the $z$-axis. However, in experiments, we cannot realize ideal instantaneous $\delta$ kicks, but only kicks of finite widths. Hence, the above equation for kicks of finite width can be expressed as
	\begin{align}
	H_{\mathrm{qkt}} = \begin{cases}
	H_{\mathrm{kick}} = \frac{\pi}{2\Delta}J_y,~{\mathrm{for~}}t\in \left[n\tau-\frac{\Delta}{2},n\tau+\frac{\Delta}{2}\right]\\
	H_{\mathrm{nl}} = \frac{k}{2j\tau}J_z^2, ~{\mathrm{otherwise.}}
	\end{cases} 
	\label{Hqkt}
	\end{align}
	Here, $\Delta$ is the kick duration that produces a $\pi/2$ rotation about the y-axis described by the unitary operator $U_{\mathrm{kick}} = \exp\{-iH_{\mathrm{kick}}\Delta\}$. The second term describes the nonlinear evolution governed by the chaoticity parameter $k$ for a time period $\tau$ with the corresponding unitary $U_{\mathrm{nl}} = \exp\{-iH_{\mathrm{nl}}\tau\}$. The effective Floquet operator can then be written as ${\cal U} = U_{\mathrm{nl}}U_{\mathrm{kick}}$. The dynamics of the system can be evaluated from the evolution of angular momentum components of the QKT under the Floquet evolution after the $n$-th kick as $J_\alpha(n+1) = {\cal U}^\dagger J_\alpha(n) {\cal U}$, for $\alpha = \{x,y,z\}$. The classical map can be obtained from the scaled variables $V = J_\alpha/j$ in the limit $j \rightarrow \infty$ \cite{haake1987classical} which leads to the following equations of motion:
	\begin{align}
	X' & = Z \cos(kX) + Y \sin(kX) \nonumber \\
	Y' & = -Z \sin(kX) + Y \cos(kX) \nonumber \\
	Z' & = -X.
	\label{classicaleq}
	\end{align}
	
\begin{figure}[t]
		\centering
		\includegraphics[trim=1.6cm 3.7cm 4cm 2.5cm,clip=,width=\columnwidth]{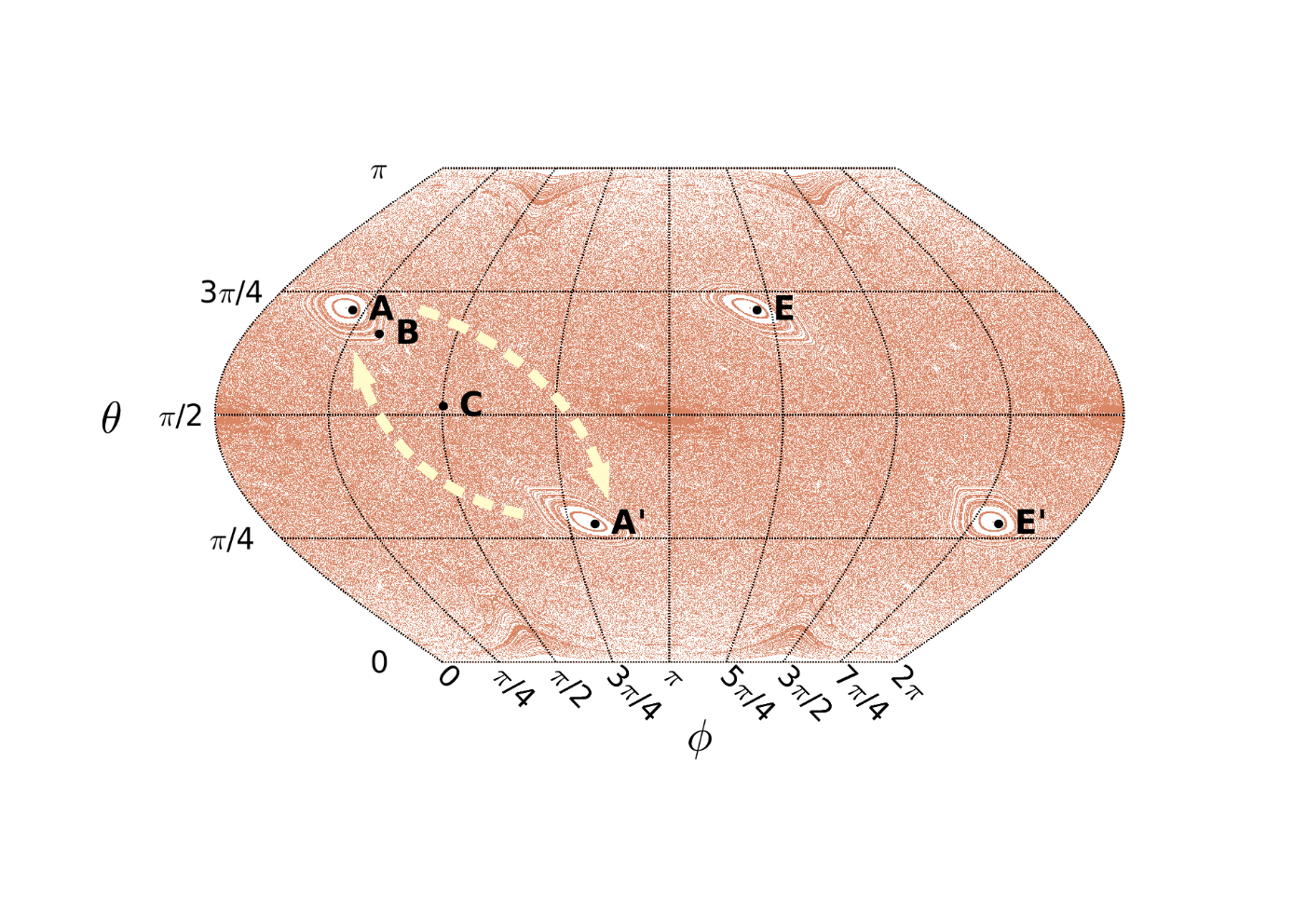}
		\caption{Classical phase space of the kicked top model for chaoticity parameter $k = 3$. The mixed phase space has distinct regular islands separated by a chaotic sea. A classical system initialized in the regular regions, labelled by \textbf{A} and \textbf{A'}, will continue to remain there throughout the dynamics, while that initialized in the near-regular region, labelled by \textbf{B}, can move along its periodic orbit, and that initialized in the  chaotic region, labelled by \textbf{C}, can explore the phase space. The regions labelled by \textbf{E} and \textbf{E'} form a period-two orbit and keep jumping from one to the other with every kick. The dynamics of a QKT initialized in the states \textbf{A}, \textbf{B}, and \textbf{C} studied here reveal dynamical tunneling between \textbf{A} and \textbf{A'} as indiated by the arrows.}
		\label{classicalphasesp}
	\end{figure}

	Since the total angular momentum of the system is conserved, the dynamics of the system can be parameterized in terms of two parameters $(\theta, \phi)$ such that $X = \sin\theta\cos\phi, Y = \sin\theta\sin\phi, Z = \cos\theta$. For low values of the chaoticity parameter, $k \sim 0.5$ the system is highly regular, but transitions to a mixed phase space as $k$ is increased before becoming almost completely chaotic at around $k = 6$ \cite{UdaySignatures}. This map has time reversal symmetry and reflection symmetry about the $y$-axis \cite{haake1987classical}. The classical phase space for $k = 3$ is shown in Fig. \ref{classicalphasesp}.  Under classical evolution, even as time $t \to \infty$, the initial conditions indicated by \textbf{A} and \textbf{A'} in Fig. \ref{classicalphasesp} will remain trapped in their respective regular regions. However, if the system is initialized in a chaotic region, indicated by \textbf{C} in Fig. \ref{classicalphasesp}, it can then explore the entire connected chaotic layer of the phase space. In contrast, a QKT initialized in one of the regular regions can tunnel to other regular regions, giving rise to dynamical tunneling, as explained below.
	
	\subsection{Dynamical tunneling in QKT model}
	%
	\begin{figure*}[t]
		\centering
		\includegraphics[trim=4cm 0.2cm 4.2cm 1cm,clip=,width=1.7\columnwidth]{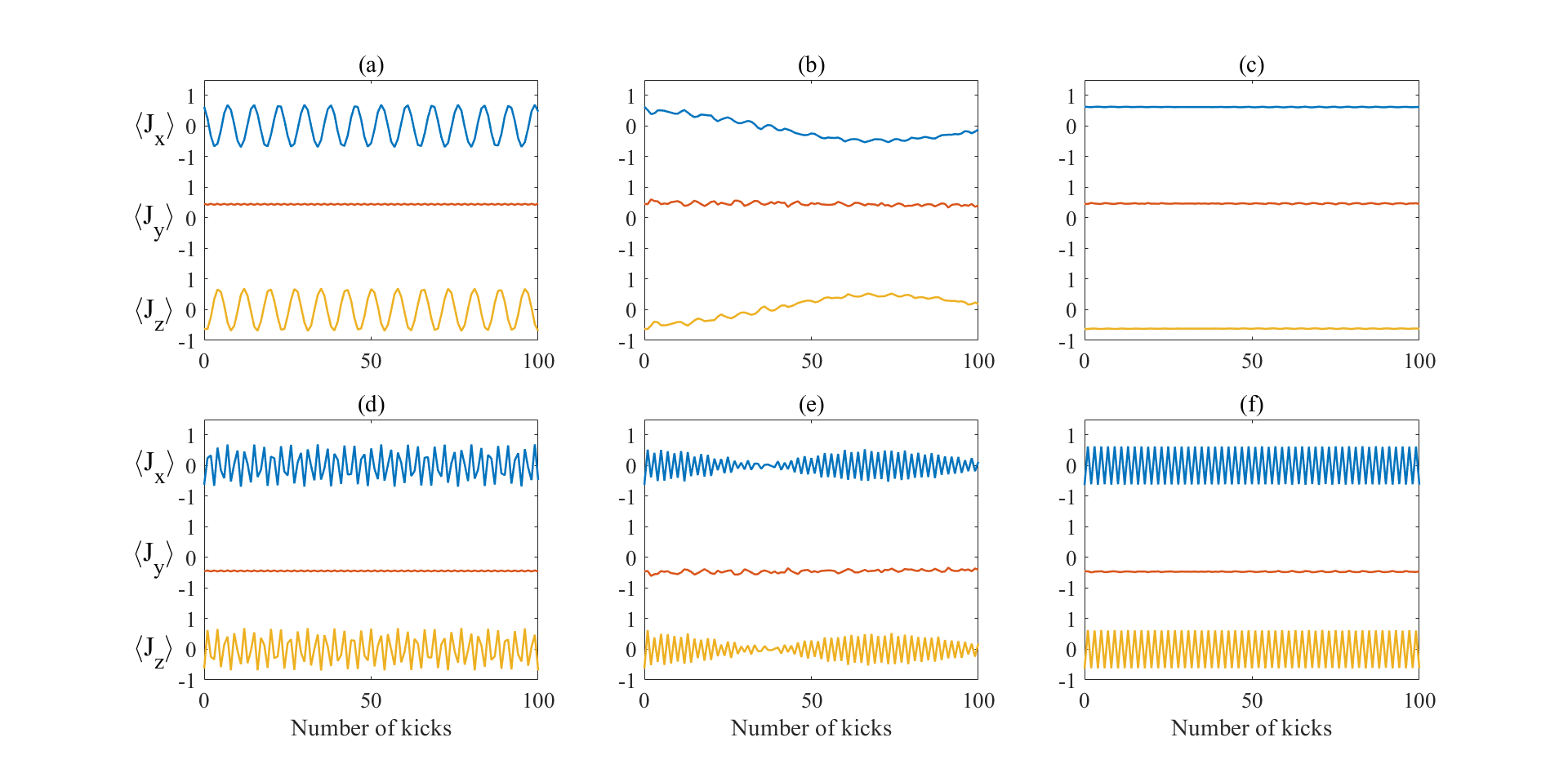}
		\caption{Normalized expectation values of angular momentum operator components $\expv{J_\alpha}$ obtained from numerical simulations with $k=3$ starting from the states \textbf{A} (a-c) and \textbf{E} (d-f) for spin sizes $j=1$ (a,d),  $j=10$ (b,e) and $j=100$ (c,f). As the spin size increases the system tends towards the classical limit exhibiting prolonged tunneling periods. For the latter initial state (d-f), the oscillations in expectation values $\expv{J_\alpha}$ are maintained for all spin sizes, with the system exhibiting clear period-two oscillations as it tends to the classical limit, which can be seen prominently for $j=100$ (f).}
		\label{spinSsim}
	\end{figure*}
	Just as a wave packet can tunnel through a potential barrier with higher energy, quantum systems can overcome \textit{dynamical barriers} and couple regular regions which are classically disconnected and between which any classical transport is strictly forbidden. A classical system initialized in one of the regular regions, \textbf{A} and \textbf{A'} in Fig. \ref{classicalphasesp} remains localized, while a quantum system can defy the classical dynamical barrier and periodically tunnel to and from the other regular region of appropriate symmetry \cite{peres}. Such periodic tunneling behaviour was theoretically studied in Ref. \cite{sanders1989effect} using the QKT model for a spin-$j=18$ system. The tunneling phenomenon was captured using the expectation values $\langle J_\alpha \rangle$ of the angular momentum operator. Periodic revivals in expectation values of $\langle J_\alpha \rangle$ indicated tunneling between regular regions, while lack of clear oscillations indicated absence of tunneling. Interestingly, a QKT initialized in a regular region showed clear periodicity, while that initialized in a chaotic region did not show such clear periodicity.
	
	While our experiments use the same QKT model, we first numerically study the system size dependence of tunneling behaviour for chaoticity parameter $k = 3$. As the system size increases ($j \to \infty$), the classical limit is approached, and the tunneling behaviour is suppressed. Let us consider the initial state $\textbf{A} \equiv \ket{\theta_\textbf{A},\phi_\textbf{A}} \equiv (2.25,0.63)$ at the centre of one of the regular regions and its symmetry related state $\textbf{A'} \equiv \exp(-i\pi J_y)\ket{\theta_\textbf{A},\phi_\textbf{A}}\equiv (\pi-2.25,\pi-0.63)$ (see Fig. \ref{classicalphasesp}).
	The numerical simulations of $\langle J_\alpha \rangle$ for the QKT model for different spin sizes starting from \textbf{A} are shown in Fig. \ref{spinSsim}(a)-(c). 
	It is clear that $\langle J_x \rangle$ and $\langle J_z \rangle$ show rapid oscillations for $j=1$ (Fig. \ref{spinSsim}(a)) indicating tunneling between \textbf{A} and \textbf{A'}. However, for a larger system with $j=10$ (Fig. \ref{spinSsim}(b)) the period is elongated, and for $j=100$ (Fig. \ref{spinSsim}(c)) the system shows no sign of periodicity in the chosen time range. It is interesting to note that the other pair of similar-looking regular regions, labelled by $\textbf{E} \equiv \ket{\theta_\textbf{E},\phi_\textbf{E}} = (2.25,0.63+\pi)$ and $\textbf{E'} \equiv \exp(-i\pi J_y)\ket{\theta_\textbf{E},\phi_\textbf{E}}\equiv (\pi-2.25,2\pi-0.63)$, have a totally different behaviour, as shown in Fig. \ref{spinSsim}(d)-(f). They form a period-two orbit and oscillate between one another with every kick in the classical limit \cite{haake1987classical}. This is clearly observed for a large spin system, such as $j=100$ in Fig. \ref{spinSsim}(f). For smaller spin sizes, such as $j=1$ and $j=10$ (Fig. \ref{spinSsim}(d,e)), the values of $\expv{J_x}$ and $\expv{J_z}$ show irregular oscillations with beat patterns.

	\section{Experimental methodology}
	\label{NMRExp}
	\subsection{NMR Hamiltonian}
	\begin{figure}[t]
		\centering
		\includegraphics[trim=2.5cm 5.2cm 5cm 3.5cm,clip=,width=\columnwidth]{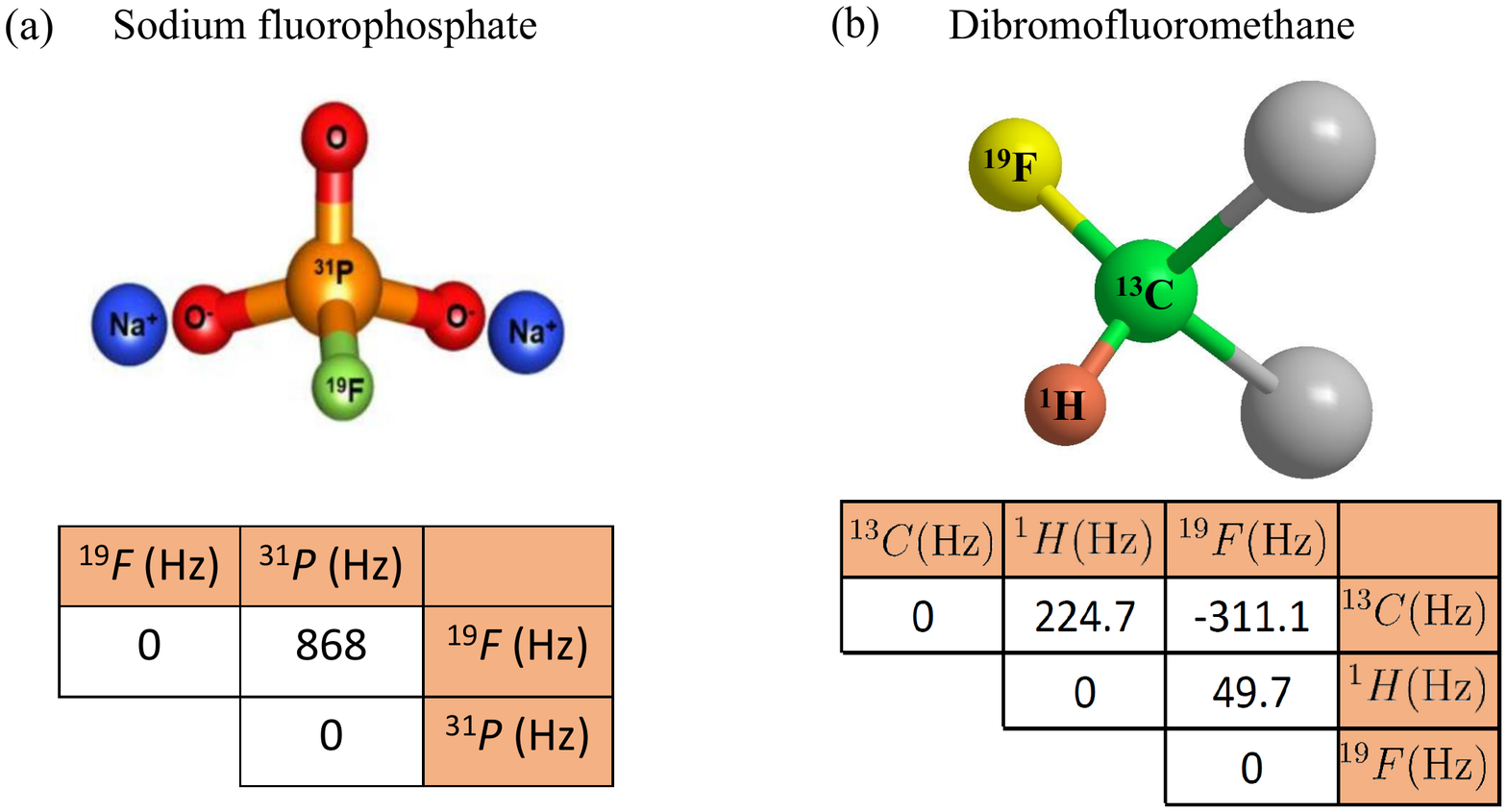}
		\caption{Experimental systems used for tunneling experiments. (a) The two-qubit system of sodium fluorophosphate used to simulate a single spin-1 system and (b) three-qubit system of dibromofluoromethane used to simualte a single spin-3/2 system, along with their Hamiltonian parameters shown in the tables below. The diagonal elements indicate chemical shifts, while off-diagonal elements indicate the scalar $\cal J$-coupling constant values.}
		\label{molecules}
	\end{figure}
	To study the size dependent behaviour of dynamical tunneling, we simulated the QKT in spin-1 and spin-3/2 systems using two- and three-qubit NMR systems respectively. The two-qubit system comprised $^{19}$F and $^{31}$P of sodium fluorophosphate (Fig. \ref{molecules}(a)) dissolved in D$_2$O, and the three-qubit system comprised $^{13}$C, $^{1}$H and $^{19}$F spins of dibromofluoromethane  (Fig. \ref{molecules}(b)) dissolved in deuterated acetone. All the experiments were performed on samples containing about $10^{15}$ nuclear spins maintained at 300 K on a Bruker 500 MHz high resolution spectrometer with a static magnetic field $B_0 \hat{z}$ with $B_0=11.7$ T. 
	The field lifts the degeneracy of $m_s$ spins levels via the Zeeman interaction, with an energy gap $\hbar\gamma_i B_0$ which has the corresponding Larmor frequency of $\omega_i = \gamma_i B_0$ where $\gamma_i$ is the gyromagnetic ratio of the spin \cite{levitt2013spin}. Different nuclear spin species exist in different chemical environments which influence the effective field experienced by the spins. The resulting time-averaged local field corresponds to a  modified Larmor frequency $\omega_i = \gamma_i B_0 (1+\delta_i)$, where $\delta_i$ is the chemical shift of the spins. The spins also interact indirectly with one another via the scalar coupling constant ${\cal J}_{ij}$ mediated by covalent bonds.  For the heteronuclear systems considered here, we move to a rotating frame resonant with the Larmor frequencies of the spins and the chemical shifts may be set to zero.  The effective NMR Hamiltonian in the weak-coupling limit is then given only by the scalar ${\cal J}_{ij}$ coupling interaction and takes the form
	\begin{equation}
	H_{\cal J} = \sum_{i,j>i} 2\pi {\cal J}_{ij} I_{zi} I_{zj}.
	\end{equation}
	The spins can further be manipulated by radio frequency (RF) pulses resonant with the corresponding characteristic Larmor frequencies and described by the Hamiltonian
	\begin{equation}
	H_{\mathrm{RF}} = \sum_i \frac{\pi}{2\Delta_i} I_{yi},
	\end{equation}
	where $\Delta_i$ is the pulse duration corresponding to the $i$-th spin species. Hence the NMR system with the RF pulses is described by the combined Hamiltonian \cite{krithika2019}
	\begin{align}
	H_{\mathrm{NMR}} & = \sum_i \frac{\pi}{2\Delta_i} I_{yi} + \sum_{i,j>i} 2\pi {\cal J}_{ij} I_{zi}I_{zj}.
	\end{align}
	In systems with three or more qubits,  we can realize an uniform evolution under a single effective scalar coupling constant ${\cal J}$ by using the standard spin echo methods \cite{cavanagh}, such that 
	\begin{align}
	H_{\mathrm{NMR}}^{\mathrm{eff}} & = H_{\mathrm{RF}} + H_{\cal J}^{\mathrm{eff}} \nonumber\\
	& = \sum_i \frac{\pi}{2\Delta_i} I_{yi} + {\cal J} \sum_{i,j>i} 2\pi  I_{zi}I_{zj}.
	\end{align}
	Comparing this with Eq. \ref{Hqkt}, we can see that the linear term $H_{\mathrm{kick}}$ can be mapped to the RF term $H_{\mathrm{RF}}$. Since we realize the spin-$j$ QKT using a collection of 2$j$ qubits \cite{Wang2004,UdaySignatures}, the nonlinear term in Eq. \ref{Hqkt} can be expanded as 
	\begin{align}
	\frac{k}{2j\tau} J_z^2 & = \frac{k}{2j\tau}\left(\sum_{i=1}^{2j} I_{zi}\right)^2\nonumber\\
	& = \frac{k}{2j\tau} \left[\sum_{i=1}^{2j} I_{zi}^2 + 2\sum_{i=1,j>i}^{2j} I_{zi}I_{zj}\right]\nonumber\\
	& = \frac{k}{2j\tau} \left[\sum_{i=1}^{2j} \frac{\mathbbm{1}}{4} + 2\sum_{i=1,j>i}^{2j} I_{zi}I_{zj}\right]\nonumber\\
	& \equiv \frac{k}{2j\tau} 2\sum_{i=1,j>i}^{2j} I_{zi}I_{zj}.
	\label{Jexp}
	\end{align}
	Thus, the nonlinear term can be mapped to the scalar $\cal J$ coupling term $H_{\cal J}^{\mathrm{eff}}$ up to the identity term which only introduces an unobservable global phase. Moreover, comparing Eq. \ref{Jexp} with $H_{\cal J}^{\mathrm{eff}}$, we can see that $k = 2j \pi {\cal J} \tau$, which enables us to vary the chaoticity parameter $k$ by tuning the duration $\tau$ of the effective ${\cal J}$ evolution. Since the duration of the RF pulse $\Delta_i\ll\tau = k/(2j\pi{\cal J})$, we ignore $H_{\cal J}^{\mathrm{eff}}$  during the RF pulse and hence decompose the Floquet evolution ${\cal U}_{\mathrm{NMR}} = U_{\cal J}U_{\mathrm{RF}}$, where $U_{\mathrm{RF}} = \mathrm{exp}(-iH_{\mathrm{RF}}\Delta)$ and $U_{\cal J} = \exp(-iH_{\cal J}^{\mathrm{eff}}\tau)$.

	\subsubsection*{Initial state preparation}
	At ambient temperatures, the thermal energy $k_B T$ of the NMR spin system is much larger than the Zeeman energy splitting $\hbar\gamma_iB_0$. Hence, an $n$-qubit system is in a highly mixed state and is given by the Boltzmann distribution \cite{levitt2013spin}
	\begin{equation}
	\rho_{\mathrm{eq}} \simeq \frac{\mathbbm{1}}{2^n} + \sum_i \epsilon_i I_{zi},
	\end{equation}
	where $\mathbbm{1}/2^n$ captures the uniform population background, and the purity factor $\epsilon_i = \hbar\gamma_i B_0/(2^n k_B T) \sim 10^{-5}$ captures the deviation from uniform population distribution. 
	
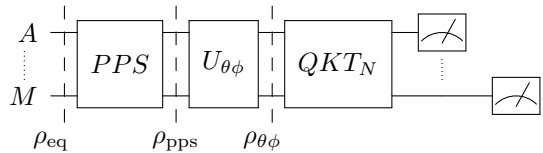
\begin{figure}[t]
		
		\hspace*{0.7cm}
		\Qcircuit @C=1em @R=1em {
			\lstick{A} \ar@{.}[]+<-1em,-0.7em>;[d]+<-1em,0.6em>  \barrier[-2.1em]{1}  & \multigate{1}{PPS} \barrier[-1.8em]{1}& \multigate{1}{U_{\theta\phi}} \barrier[-2.5em]{1} & \multigate{1}{QKT_N} & \meter \ar@{.}[]+<0em,-1em>;[d]+<0em,0.6em> &
			\\
			\lstick{M} & \ghost{PPS} & \ghost{U_{\theta\phi}} & \ghost{QKT_N} & \qw & \meter\\ 
			\rho_\mathrm{eq} & ~~~~~~~~~~~~~~~~~~~~~~\rho_\mathrm{pps} ~~~~~\rho_{\theta\phi} & & & &
		} \\ ~ \\
		\caption{An experimental circuit to realize the QKT model in a system of $M$-qubits. Starting from the state $\rho_\mathrm{eq}$ in thermal equilibrium, a pseudopure state $\rho_\mathrm{pps}$ is prepared. This is followed by preparation of initial state $\rho_{\theta\phi}$. We then implement the  QKT model for $N$ kicks and finally readout each qubit.}
		\label{circuitdiag}
	\end{figure}

	To simulate the dynamics of a QKT, it is conventional to initialize the system into coherent states as these are closest to a classical state \cite{Radcliffe1971,KlauderCoherent}. We simulate a spin-$j$ QKT using 2$j$ qubits initialized in the spin coherent state defined as 
	\begin{equation}
	\ket{\theta,\phi} = U_{\theta\phi}\ket{0}^{\otimes n}, ~\mathrm{where~} U_{\theta\phi}=e^{-i\phi \sum_iI_{zi}}e^{-i\theta \sum_iI_{yi}}.
	\end{equation}
	To realize this in a multi-qubit NMR spin system, we first transform the thermal equilibrium state $\rho_{\mathrm{eq}}$ to a pseudo-pure state of the form $\rho_{\mathrm{pps}} = (1-\epsilon) \mathbbm{1}/2^n + \epsilon \ket{\psi}\bra{\psi}$ whose dynamics can be mapped isomorphically to the dynamics of a pure state $\ket{\psi}$  \cite{CoryPPS,gershenfeld1997bulk}. The detailed NMR pulse sequences for preparing PPS of the two- and three-qubit spin systems considered here are given in \cite{krithikaBlockade}. These states can then further be transformed into coherent states $\ket{\theta,\phi}$ for an $n$-qubit system 
	\begin{align}
	\rho_{\mathrm{\theta\phi}} & = U_{\theta\phi}\rho_{\mathrm{pps}}U_{\theta\phi}^\dagger \equiv \ket{\theta,\phi}\bra{\theta,\phi}.
	\end{align}
	The system is thus initialized to a required $(\theta,\phi)$ coordinate in the phase space and the QKT Floquet operator ${\cal F}_{\mathrm{NMR}}$ is subsequently applied $N$ times to study the time evolution. An experimental circuit, showing the line-up of successive operations for simulating a QKT is displayed in Fig. \ref{circuitdiag}.
	
	\subsubsection*{Measurement of $\expv{J_\alpha}$}
	In an NMR system, the direct signal measurement by quadrature detection gives $\expv{I_{xi}} + i\expv{I_{yi}}$ \cite{levitt2013spin}. 
	To extract $\expv{I_{zi}}$, we apply the following in succession: ({\it i}) a pulsed field gradient (PFG) which destroys the $x$ and $y$ magnetization components of the system and following this, ({\it ii}) a ($\pi$/2) pulse about the $y$-axis to rotate the $z$-component of magnetization to the $x$-axis, and then detect the transverse magnetization.  Note that measurement of the angular momentum components of individual spins suffices to estimate the total expectation values $\expv{J_{\alpha}}$ (see Appendix B).
	In the following, we discuss the results of the above mentioned protocols for studying dynamical tunneling in two- and three-qubit spin systems.

	\section{Experimental Results}
	\label{Res}
	\subsubsection{Tunneling in mixed phase space ($k = 3$)}
	\begin{figure}
		\centering
		\includegraphics[trim=0.5cm 0.1cm 0.5cm 0cm,clip=,width=\columnwidth]{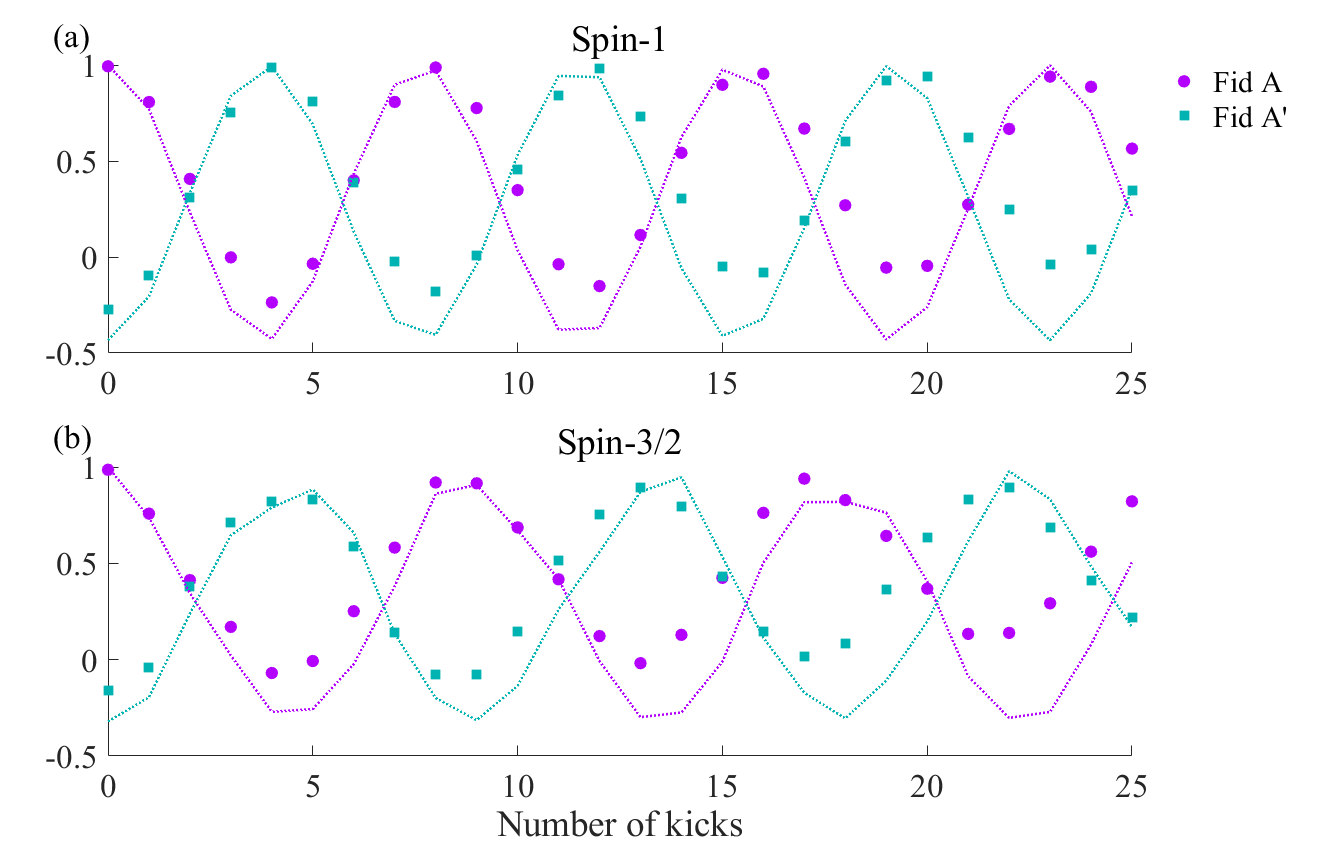}
		\caption{Trace fidelity of the instantaneous state of the system initialized in the regular region \textbf{A} for spin-1 (a) and spin-3/2 (b) systems with respect to the tunneling regions \textbf{A} and \textbf{A'}. Experimentally extracted values of fidelity are indicated by symbols overlaid on simulated values indicated by dotted lines. }
		\label{fid2q3q}
	\end{figure}
	
	As explained above, we initialize the two- and three-qubit based QKT systems to different regions of the mixed phase space at $k = 3$, and study the tunneling behaviour via $\expv{J_\alpha}$ for $\alpha \in [x,y,z]$. Following Sanders and  Milburn's work (Ref. \cite{sanders1989effect}) we chose the initial state $\textbf{A}$ (see Fig. \ref{classicalphasesp}) in the regular region of phase space, while the initial state $\textbf{B}$ lies in the border between regular region and chaotic sea. The initial state $\textbf{C}$ lies entirely in the chaotic sea. The system was evolved for $N=25$ kicks and $\expv{J_\alpha}$ was measured after each kick. Note that a classical system initialized in state \textbf{A} in the regular region is dynamically bound and cannot escape to other regions, such as the state \textbf{A'}.
	
	\begin{figure*}
		\centering
		\includegraphics[trim=2.7cm 5.5cm 4cm 1cm,clip=,width=17.5cm]{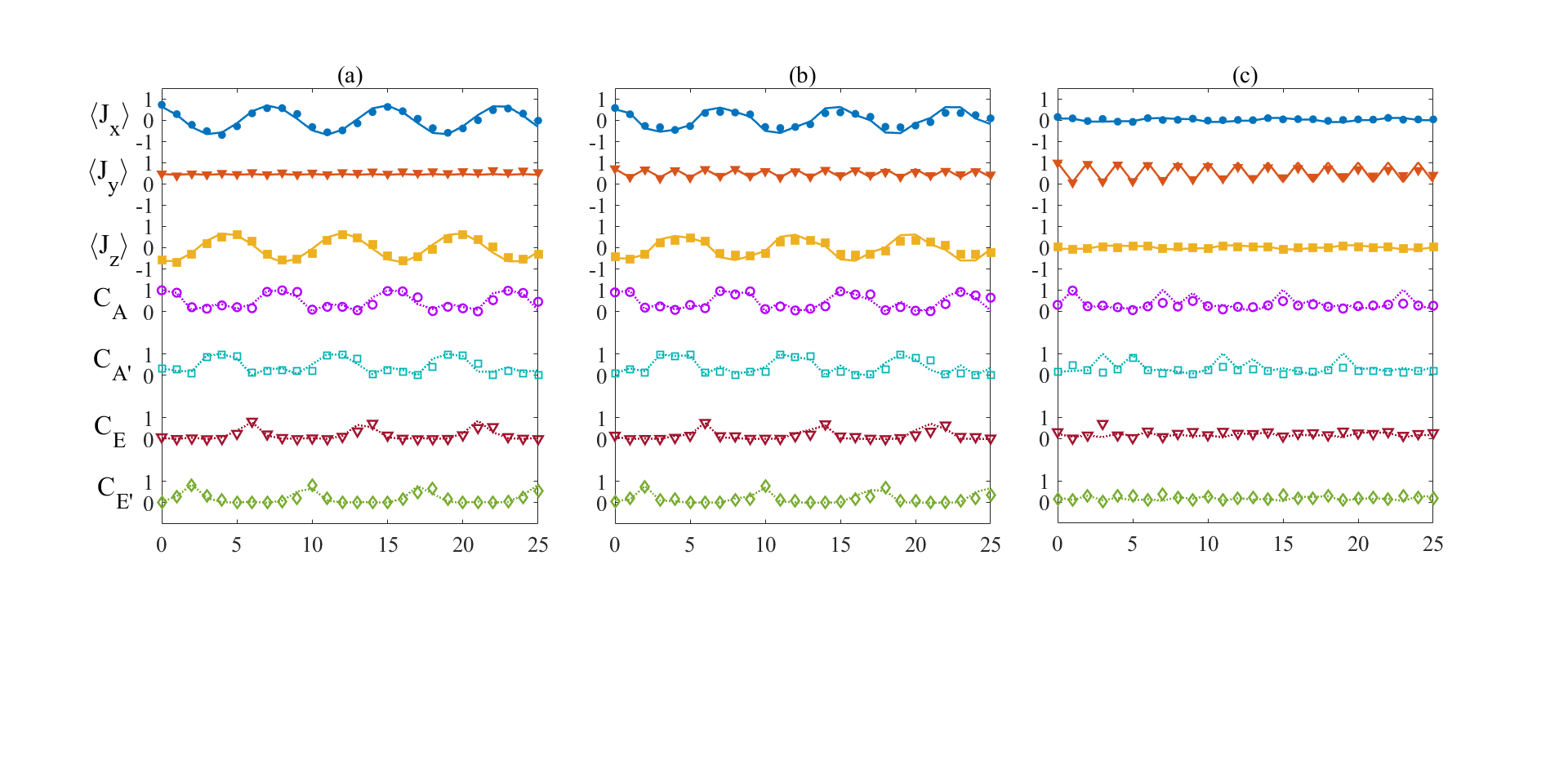}\\
		\includegraphics[trim=2.7cm 5.5cm 4cm 2cm,clip=,width=17.5cm]{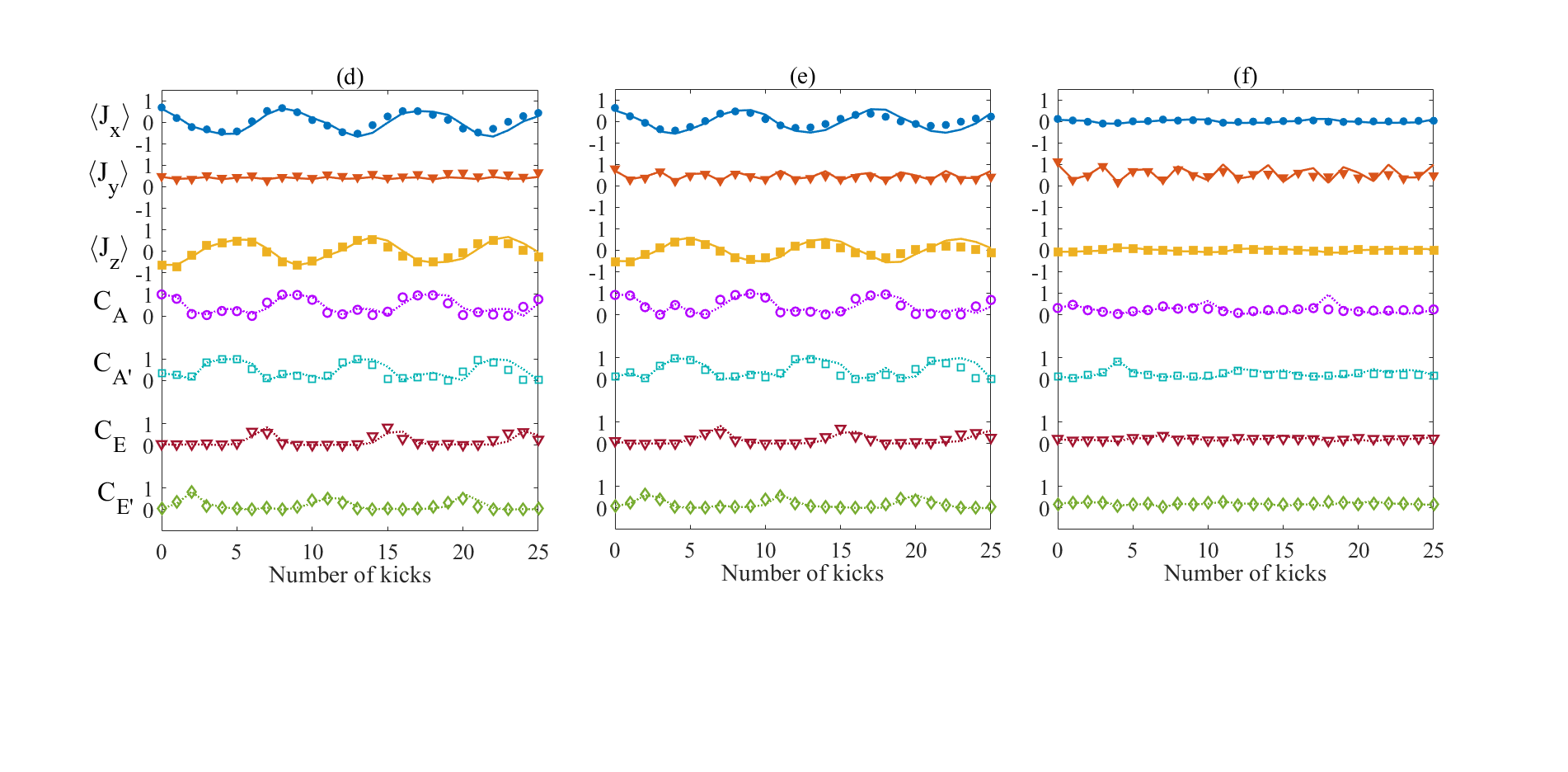}
		\caption{Dynamics of QKT in two-qubit spin-1 system (a-c) and three-qubit spin-3/2 system (d-f) corresponding to initialization in states \textbf{A} (a,d), \textbf{B} (b,e), and \textbf{C} (c,f). The symbols indicate experimental data while dashed lines indicate simulation.  The upper three traces represent $\expv{J_\alpha(t)}$ and the lower four traces represent $C_S(t)$.  In (a,d) we  see that both the systems show clear tunneling patterns for initialization in the regular region with good agreement between simulation and experiments. The revival patterns are observed for the near-regular region as well (b,e), but are not as prominent as those of the regular region. The patterns for chaotic initial state (c,f) show no clear periodicity. }
		\label{results2q3q}
	\end{figure*}
	
	\begin{figure*}
		\centering
		\includegraphics[trim=2.7cm 5.5cm 4cm 1cm,clip=,width=17.5cm]{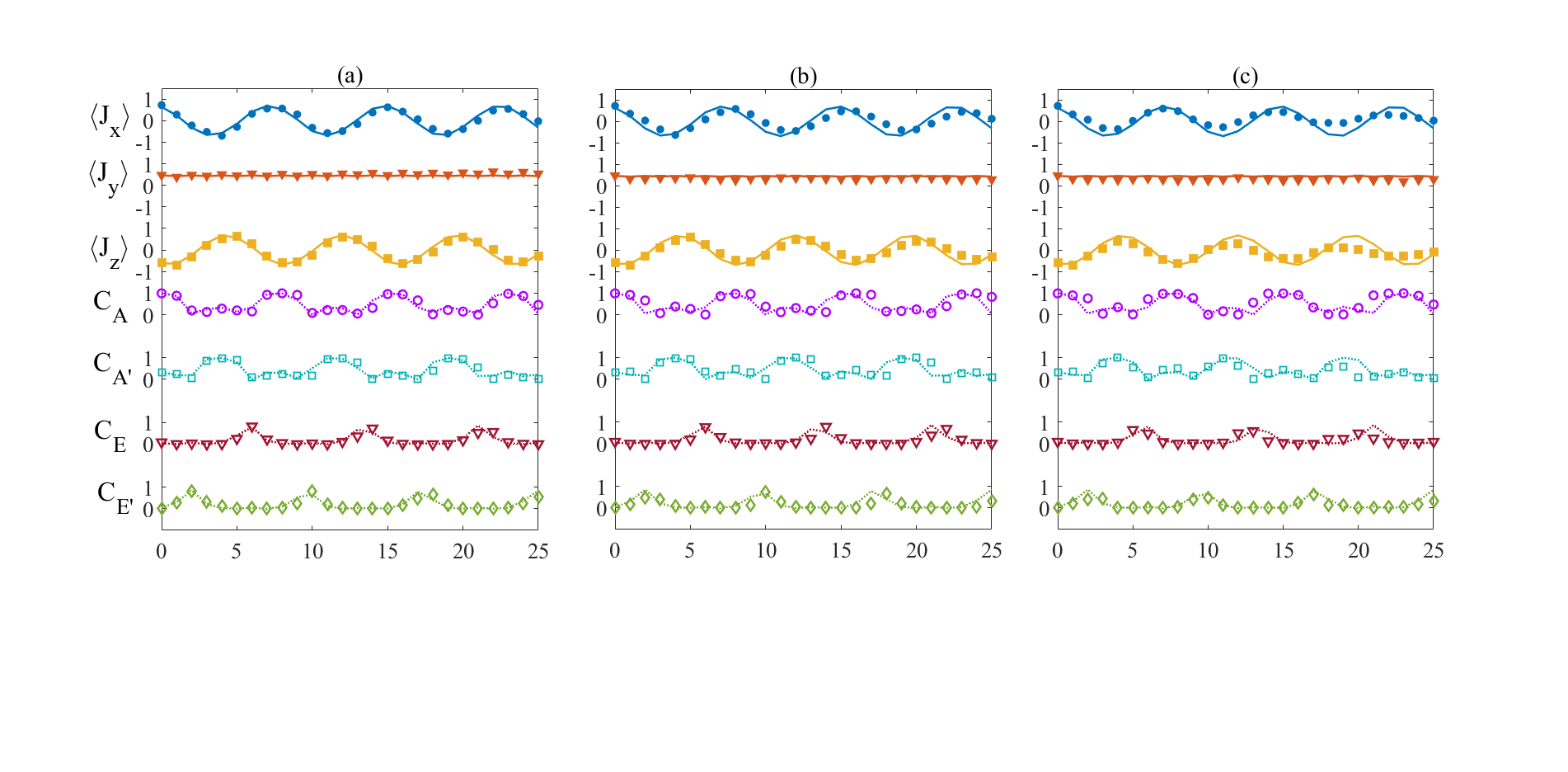}\\
		\includegraphics[trim=2.7cm 5.5cm 4cm 2cm,clip=,width=17.5cm]{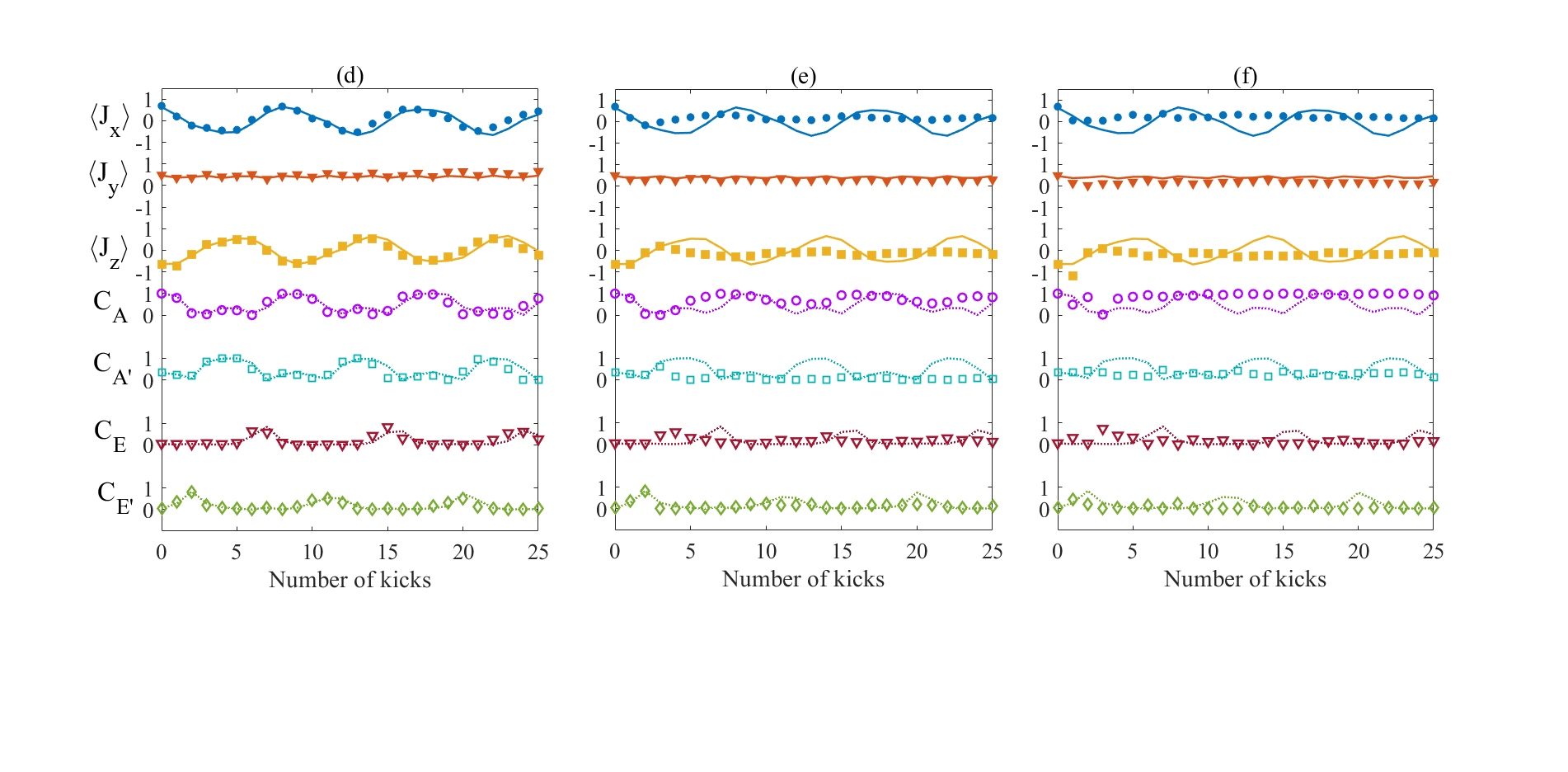}
		\caption{Effects of dephasing noise on dynamical tunneling in spin-1 QKT realized with two-qubits (a-c) as well as spin-3/2 QKT realized with three-qubits (d-f), with PFG strengths 0 G/cm (a,d), 0.005 G/cm (b,e), and 0.05 G/cm (c,f). In all the cases, the system was initialized in state \textbf{A} inside a regular region of Fig. \ref{classicalphasesp}.  The symbols indicate experimental data overlaid on dashed lines corresponding to ideal simulations without any dephasing noise.  The upper three traces represent $\expv{J_\alpha(t)}$ and the lower four traces represent $C_S(t)$. While both the systems are susceptible to dephasing noise the $j=1$ system is relatively more robust in comparison to $j=3/2$ system wherein the oscillations have decayed more severely with noise.}
		\label{diff2q3q}
	\end{figure*}
	
	When working with such small quantum systems, the spreading of wavefunctions (outside the phase space region of interest) might be significant and hence needs to be monitored to ensure that tunneling we observe is not due to leakage of probability density. To quantify the overlap of the the time-evolving state with the initial coherent state in regular region \textbf{A} and the symmetry-related tunneling region \textbf{A'}, we study the trace fidelity defined as \cite{Fortunato}
		\begin{equation}
		F_\textbf{S}(t) = \frac{\mathrm{tr}(\rho(t) \rho_\textbf{S})}{\sqrt{\mathrm{tr}(\rho(t)^2)\mathrm{tr}(\rho_\textbf{S}^2)}},
		\label{fid}
		\end{equation}
		where $\rho(t)$ is the traceless deviation density matrix of the instantaneous state of the system at time $t$, $\rho_S$ for $S\in \{\textbf{A},\textbf{A'}\}$ are the deviation density matrices of coherent states \textbf{A} and \textbf{A'}. The experimentally measured (symbols) and theoretically estimated (dotted lines) trace fidelity of systems evolving under QKT dynamics with initial state \textbf{A} are shown in Fig. \ref{fid2q3q} for spin-1 (a) and spin-3/2 (b) systems respectively. Note that the trace fidelity can take negative values since the numerator in Eq. \ref{fid} is the product of two traceless matrices. Exact overlap is quantified by $F_\textbf{S}(t) = 1$, while orthogonality is quantified by $F_\textbf{S}(t)=0$. Non-zero negative values indicate partial overlap and opposite phases between states. From Fig. \ref{fid2q3q}(a-b), it is evident that the initial coherent state has maximum overlap with the regular region \textbf{A} and a modest overlap with \textbf{A'} in spin-1 and even smaller overlap in spin-3/2. Moreover, as the system evolves under QKT dynamics, it periodically localizes in \textbf{A} and \textbf{A'} with fidelity $>0.94$ in spin-1 system and $>0.83$ in spin-3/2 system. 
	
	Fig. \ref{results2q3q} shows the experimental results (symbols) and numerical simulations (dotted lines) for spin-1 system realized using two-qubits (Fig. \ref{results2q3q}(a-c)) and spin-3/2 system realized using three-qubits (Fig. \ref{results2q3q}(d-f)) initialized in states $\textbf{Q}\in \{\textbf{A},\textbf{B},\textbf{C}\}$ of the classical phase space shown in Fig. \ref{classicalphasesp}. 
	In all cases, we set the chaoticity parameter $k = 3$ and initialized the systems in states \textbf{A} (Fig. \ref{results2q3q}(a,d)), \textbf{B} (Fig. \ref{results2q3q}(b,d)), and \textbf{C} (Fig. \ref{results2q3q}(c,f)). In all the graphs, the top three traces show the expectation values $\expv{J_\alpha^\textbf{Q}(t)}$.
	For the initialization into state \textbf{A} in the regular region, we observe prominent oscillations in the expectation values of $J_x$ and $J_z$, while that of $J_y$ remains constant as the system is symmetric about $y$-kicks (see Fig. \ref{results2q3q}(a,d)). A state initialized in \textbf{B} near the border of regular and chaotic region shows similar periodicity, though not as prominent as that for \textbf{A} (see Fig. \ref{results2q3q}(b,e)). For initial state \textbf{C} in the chaotic region, we observe no clear periodicity, although the $J_y$ component shows oscillation as the system periodically gets localized and delocalized with kicks  (see Fig. \ref{results2q3q}(c,f)). The experimental data shows a decay in the oscillations due to decoherence and other experimental imperfections. We note that relatively longer time period of three-qubit oscillations compared to that of the two-qubit system (see Appendix D for further analysis).
	
	In all the plots, the lowest four traces show correlations
	\begin{align}
	C_{\textbf{S}}(t) & = \vert\langle J^{\textbf{S}}\vert J^{\textbf{Q}}(t)\rangle\vert^2 
	\end{align}
	between $J^{\textbf{S}}$ of state $\textbf{S}$ and the instantaneous total angular momentum operators $J^{\textbf{Q}}(t)$. The overlap measure allows us to track the localization of the system in states \textbf{A} and \textbf{A'} as it tunnels between these regular regions. As expected, when the system is initialized in state \textbf{A}, we see clear periodic and out-of-phase tunneling oscillations of $C_{\textbf{A}(\textbf{A'})}(t)$ (see Fig. \ref{results2q3q}(a,d)). These tunneling oscillations persist even for near-regular initialization in state \textbf{B} due to significant spreading of the low dimensional quantum systems considered here (see Fig. \ref{results2q3q}(b,e)). However, such tunneling oscillations are washed out for chaotic initialization in state \textbf{C} ((see Fig. \ref{results2q3q}(c,f)). Furthermore, the correlation measures $C_\textbf{S}(t)$ indicate that for the chaotic state, it is widely delocalized.  The bottom two traces in Fig. \ref{results2q3q} capture brief leakage amplitudes to the regions \textbf{E} and \textbf{E'}, which is the consequence of deep-quantum systems considered here.

	
	\subsubsection{Robustness of dynamical tunneling}
	Now that we observe tunneling across a dynamical barrier, it is interesting to see the role of quantum coherence in sustaining tunneling. To this end, we monitor the robustness of dynamical tunneling between regular regions \textbf{A} and \textbf{A'} under dephasing noise. For this purpose, we use pulsed field gradients (PFG) which introduce a linearly varying magnetic field along the $z$-direction and accordingly distributing Larmor frequencies over the length of the sample \cite{cavanagh}. PFG along with translational diffusion of molecules, effectively induces strong dephasing in the system. The experimental impact of dephasing on dynamical tunneling are shown in Fig. \ref{diff2q3q} for $j = 1$ (Fig. \ref{diff2q3q}(a-c)) and $j = 3/2$ systems (Fig. \ref{diff2q3q}(d-f)) and for PFG strengths $0$ G/cm (Fig. \ref{diff2q3q}(a,d)), $0.005 $ G/cm (Fig. \ref{diff2q3q}(b,e)), and $0.05$ G/cm (Fig. \ref{diff2q3q}(c,f)). The $0$ G/cm scenario in Fig. \ref{diff2q3q}(a,d) is the same as Fig. \ref{results2q3q}(a,d) and has been replotted here for visual comparison. For reference, we have plotted the theoretical lines in Fig. \ref{diff2q3q} without any dephasing effects. 
	We find that in both $j=1$ and $j=3/2$ cases, the tunneling behaviour is weakened by dephasing noise. In the two-qubit system, the periodic oscillations survive, but with decaying tunneling amplitudes (see Fig. \ref{diff2q3q}(b,c)). In the three-qubit case, even in the presence of weak PFG of 0.005 G/cm the oscillations decay much faster (see Fig. \ref{diff2q3q}(e,f)). Here, the correlation measure indicates that the system preferentially larger overlap with the regular region \textbf{A} compared to other regular regions.
	These results indicate the fragility of dynamical tunneling under dephasing noise, and thereby establish the importance of quantum coherence in sustaining the phenomenon.
	
	\section{Conclusions}
	\label{Conc}
	Dynamical tunneling, such as the chaos-assisted tunneling, is a well studied phenomenon and has been demonstrated experimentally in driven cold atomic cloud, microwave annular billiard and has most recently been used to generate NOON states \cite{VanhaeleNOON2022}. However, a systematic study of tunneling with system size and different initial conditions was not available. In this work, we have experimentally demonstrated chaos-assisted tunneling in two- and three-qubit systems using NMR based test bed. 
	We initialized the systems to different regions of the phase space -- regular, near-regular (border region between regular and chaotic) and chaotic. Following \cite{sanders1989effect}, we use $\expv{J_\alpha}$, the components of the angular momentum operator, as probes to study dynamical tunneling. We observe that the systems initialized in the regular region show periodic oscillation in $\expv{J_\alpha}$. Systems initialized in the near-regular also show periodicity in $\expv{J_\alpha}$, but the oscillations are not as perfect as those for the case of initial state in a regular region. Further, systems initialized in a chaotic region show no periodicity. Additionally, by analyzing the norm-distance between the instantaneous total angular momentum operator and that corresponding to either of regular regions, we monitor the periodic tunneling of the system between these regions for different initial conditions.
	
	To understand the significance of quantum coherence in maintaining dynamical tunneling, we studied the robustness of tunneling against dephasing noise. Experimental results showed that while both the spin $j=1$ and $j=3/2$ systems are susceptible to dephasing noise, the effect was severe for the larger system, wherein the revivals of $\expv{J_\alpha}$ were almost completely destroyed in the presence of dephasing noise.
	
	Tunneling suppression for increasing number of qubits will be related to the $\hbar$-scaling in the kicked top model. For the  QKT, quantum correlations are known to decay in a power-law form as a function of $\hbar$ \cite{UdaySignatures}. It will be useful to explore the validity of this prediction for dynamical tunneling in future studies. This is likely to be a challenging exercise from an experimental point of view since it will require maintaining coherence with large number of interacting spins. Further, while it might not be entirely surprising that introduction of noise kills tunneling effects, there are Floquet engineering techniques that allow calibrated disorder while still suppressing decoherence \cite{FloquetEngg1,FloquetEngg2}. It will be interesting to explore if such Floquet schemes help sustain chaos-assisted tunneling even in the presence of noise.  Another interesting topic to consider would be a scenario of quantum tunneling in the simultaneous presence of a potential-energy barrier as well as a dynamical barrier.   These aspects will be considered in a later work.
	
	\section{Acknowledgments}
	The authors acknowledge valuable discussions with Conan  Alexander and Prof. Arul Lakshminarayan.   We thank Dr. S. Aravinda for his inputs on improving the manuscript. M.S.S.  acknowledges  the  MATRICS  Grant No.  MTR/2019/001111  from  SERB,  DST,  Government  of India. T.S.M.  acknowledges  funding  from DST/ICPS/QuST/2019/Q67. We thank the National Mission on Interdisciplinary Cyber Physical Systems for funding from the DST, Government of India through the I-HUB Quantum Technology Foundation, IISER-Pune.  
	
	\section*{Appendix}
	\subsection{Wavefunction spreading in deep-quantum limit}
\begin{figure}[h]
		\centering
		\includegraphics[trim=0.2cm 0.1cm 0.1cm 0cm,clip=,width=\columnwidth]{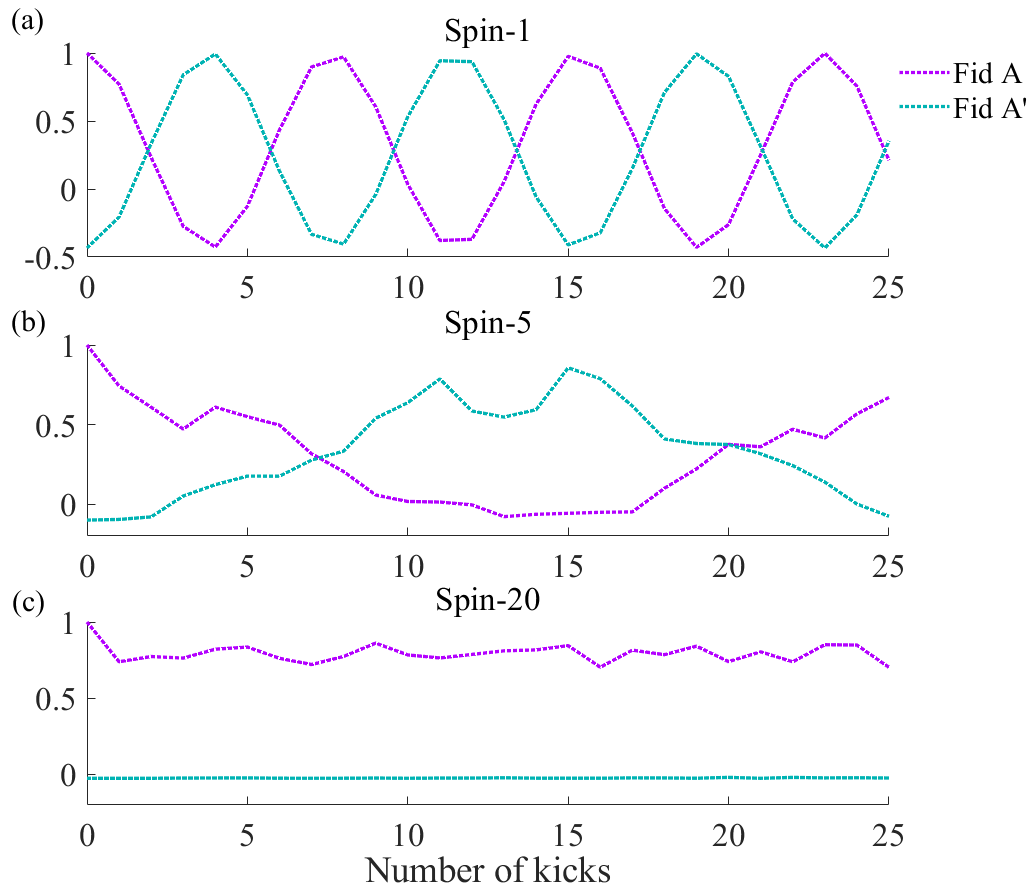}\\
		\includegraphics[trim=0cm 0cm 0cm 0cm,clip=,width=0.9\columnwidth]{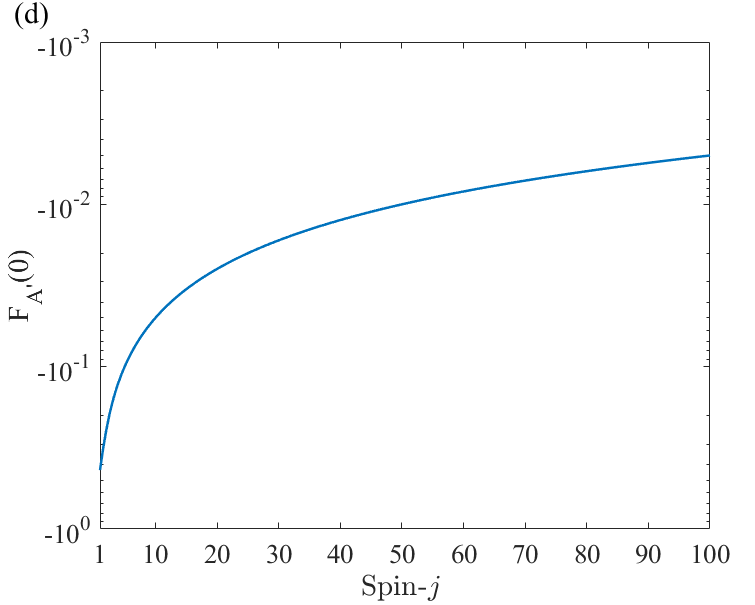}
		\caption{Trace fidelity of the instantaneous state of QKT with $k=3$ initialized in the regular region \textbf{A} for spin-1 (a) and spin-5 (b) and spin-20 (c) systems with respect to the tunneling regions \textbf{A} and \textbf{A'}. Fidelity of coherent state \textbf{A} with \textbf{A'} as a function of spin-$j$ size (d).}
		\label{fid2q3qS}
	\end{figure}

	In the deep-quantum limit, it is important to take into account the spread or finite width of spin wavefunctions. The spin coherent state into which the system is initialized has a finite spread depending on the spin size, which for smaller spins is more than that of a larger spin. Let us now look at the extent of overlap between states localized in regions \textbf{A} and \textbf{A'}. Fig. \ref{fid2q3qS} shows the theoretical trace fidelity computed using Eq. \ref{fid} of the instantaneous state of a system initialized in \textbf{A} and undergoing QKT dynamics for $k=3$   for spin-1 (a), spin-5 (b) and spin-20 (c) systems. It is evident that as the system size increases, the degree of overlap of states localized in \textbf{A}(\textbf{A'}) with \textbf{A'}(\textbf{A}) decreases. This behaviour also emphasises the importance of chaotic states in dynamical tunneling. As the system size increases, the overlap of a localized state in a regular region (\textbf{A},\textbf{A'}) with the surrounding chaotic state decreases, which in turn hampers the tunneling efficiency as is reflected in the prolonged time periods in Fig. \ref{fid2q3qS}(b,c). The fidelity of a single spin-$j$ system in coherent state \textbf{A} with the corresponding state \textbf{A'} as a function of spin size is shown in Fig. \ref{fid2q3qS}(d). It can be seen that to achieve overlap $< 0.1$ between \textbf{A} and \textbf{A'}, we need at least spin-5, i.e., ten qubits, while overlap $< 0.01$ requires at least spin-50 (or hundred qubits), which is beyond the reach of current state of the art quantum simulators.
	
	
	\subsection{Measurement of expectation values $\expv{J_\alpha}$}
	\label{appA}
	The general state $\rho$ of the multi-qubit system  can be expanded in the product operator basis of constituent spins as 
	\begin{align}
	\rho & = \frac{\mathbbm{1}}{2^n} + \sum_i c_{\alpha i} I_{\alpha i} + \sum_{ij\alpha\beta } c_{\alpha\beta ij} I_{\alpha i} I_{\beta j} + \cdots,
	\end{align}
	where higher order spin correlation terms are not shown. The total expectation value $\expv{J_\alpha}$ for the linear term can then be estimated as 
	\begin{align}
	\expv{J_\alpha} & = \mathrm{Tr}\left[\rho \sum_i I_{\alpha i} \right] = \sum_i c_{\alpha i} I_{\alpha i}
	= \sum_i \mathrm{Tr}\left[\rho_i I_{\alpha i}\right],
	\end{align}
	where $\rho_i  = \frac{\mathbbm{1}}{2} + \sum_i c_{\alpha i} I_{\alpha i}$ are the reduced density matrices of the constituent spin systems.

	\subsection{$k=0$ control experiments}
 \begin{figure}[h]
		\centering
		\includegraphics[trim=0cm 0cm 0cm 0cm,clip=,width=2.5cm]{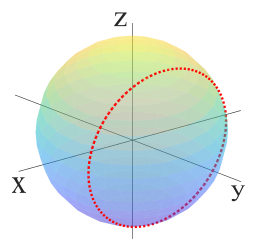} \\
		\includegraphics[trim=3cm 2.1cm 4cm 0cm,clip=,width=9cm]{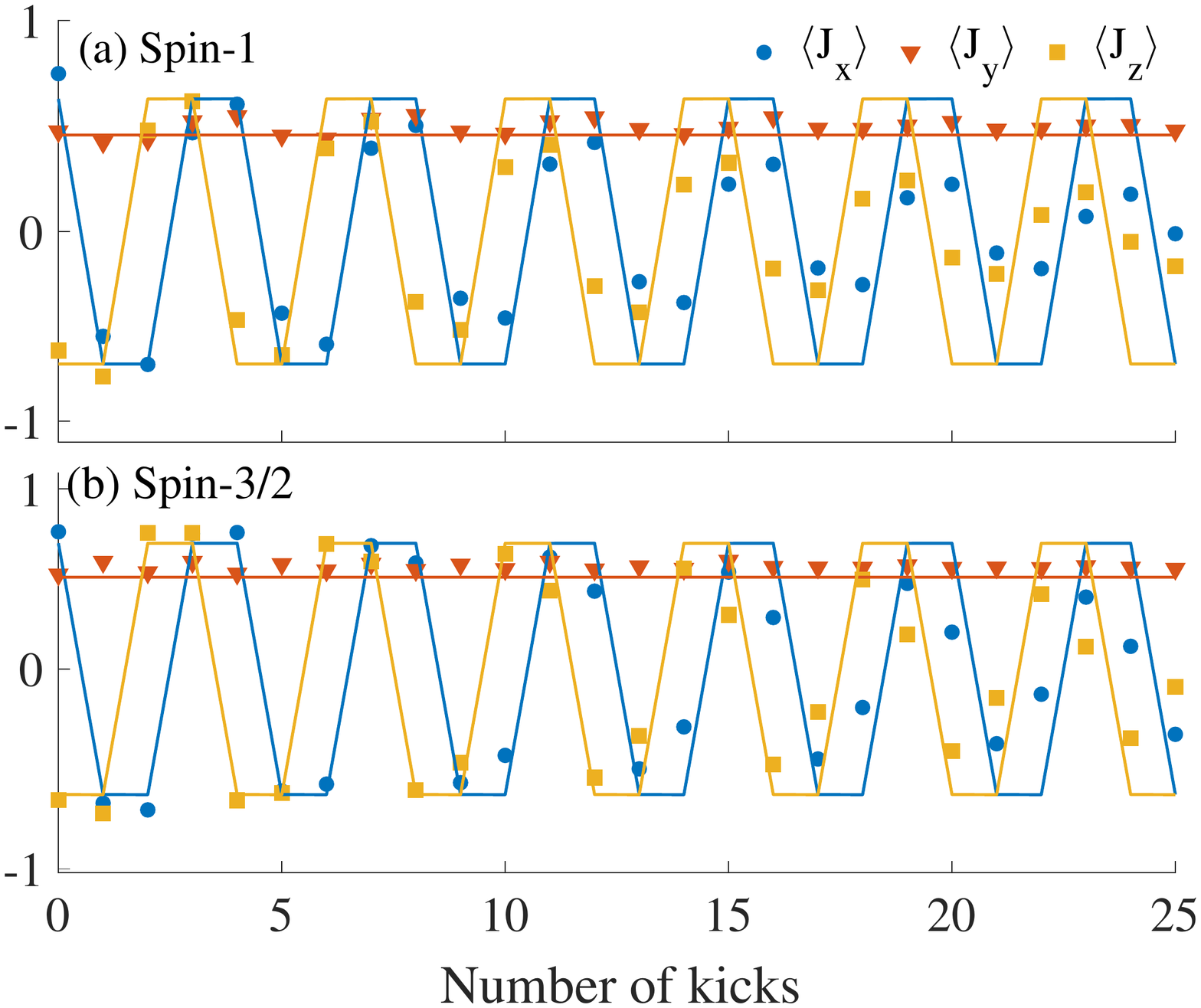} 
		\caption{Control experiments with $k=0$ for the two- and three-qubit systems. (a) Denotes the classical trajectory for different intial states. (b) Shows the data fro two- and three-qubit systems respectively. The symbols indicate experimental data, while dashed lines indicate simulations. We can see that the experimental data is in good agreement with simulated data. The decay in experimental data points is due to relaxation in the systems.}
		\label{controlexp}
	\end{figure}
	As a control, we first studied the behaviour of the system in the absence of chaos, {\it i.e.}, $k=0$. In this case, the system just evolves under $(\pi/2)$ kicks applied about the $y$-axis. The classical equations of motion (Eq. \ref{classicaleq}) at $(N+1)$-th kick relate to the $N$-th kick as follows : 
	\begin{align}
	X(N+1) & = Z(N) \nonumber\\
	Y(N+1) & = Y(N) \nonumber\\
	Z(N+1) & = -X(N).
	\end{align}
	
	The $y$-component of the system remains invariant under evolution, while the $x$ and $z$ components evolve with each kick. The evolution is thus restricted to circles in the $xz$ plane for any given initial state. The results of this control experiment are displayed in Fig. \ref{controlexp} for the system initialized in to the phase space region characterized by $\ket{\theta,\phi} = (2.25,0.63)$.
	
	\begin{figure}[h!]
		\centering
		\includegraphics[trim=1cm 0.5cm 0cm 1cm,clip=,width=8cm]{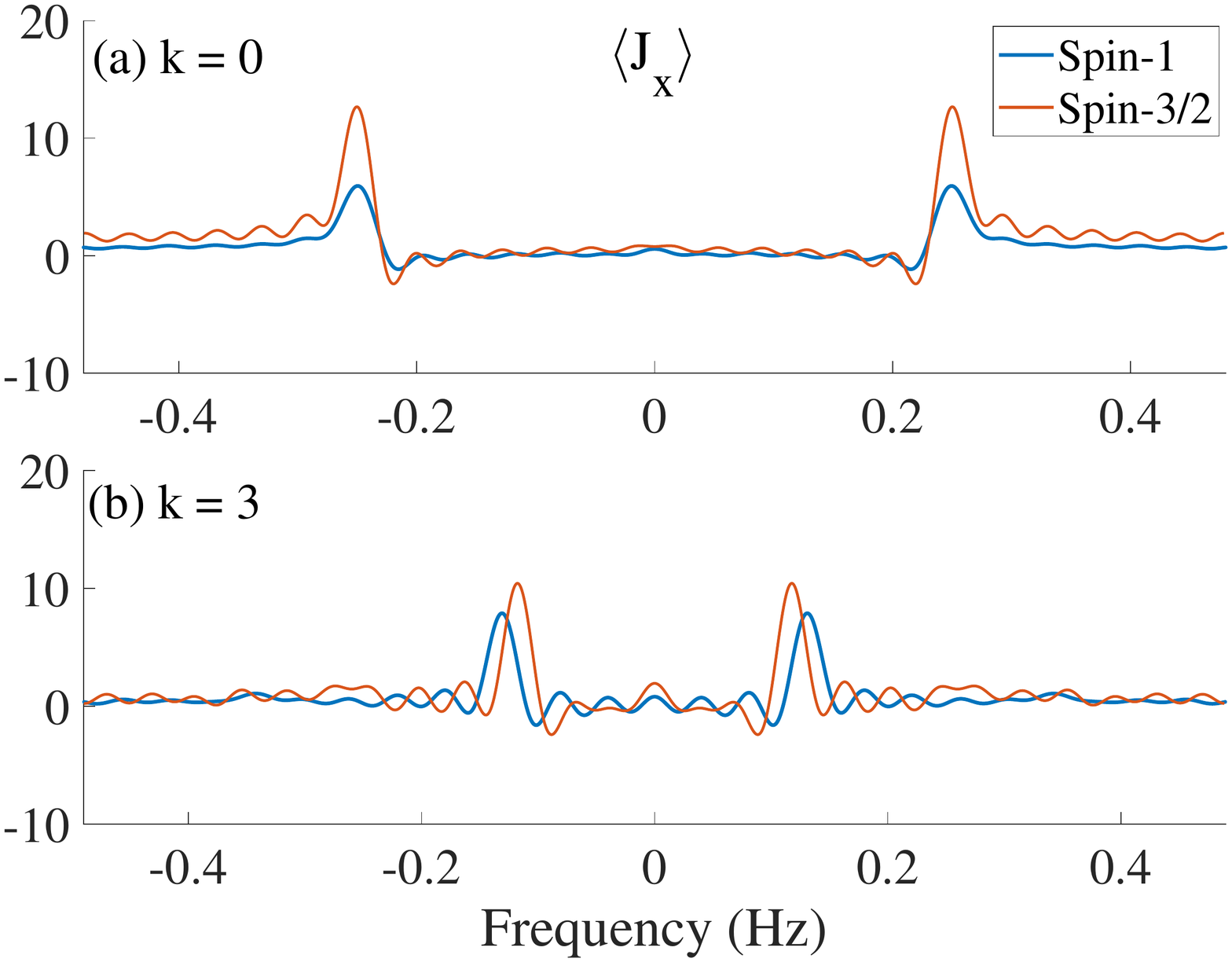}
		\caption{Fourier transform of (a) control experiments (b) tunneling experiments for spin-1 and spin-3/2 systems. We can see that in the case of control experiments, the frequency of oscillation is same for both the systems. In the case of tunneling experiments, there is a clear shift in the frequency of the three-qubit system as compared to the two-qubit system. This is in accordance with the expectation that as system size increases the tunneling effect should get suppressed.}
		\label{fft}
	\end{figure}
	The experimental data shows a decay in the amplitude of the oscillation due to accumulation of pulse errors with each kick. We can see that both the two- and three-qubit systems have oscillating ${J_x,J_z}$ values, while the value of $J_y$ remains constant. Moreover, the period of oscillation is same in both cases. To understand the frequency of oscillations better, we computed the Fourier transform of the time evolution of the system. The frequency domain analysis of the evolution (displayed in Fig. \ref{fft}(a)) shows that the period of oscillation, as anticipated, is independent of the system size. 
	
	\subsection{Time period of oscillations and system size dependence}
	Comparing the periodicity of oscillation, we can see that the period  slightly longer for the three-qubit system which completes about three oscillations in 25 kicks, while the two-qubit system completes three and a half oscillations in the same duration.  In the case of $k = 3$, the period of oscillations decreases with increasing system size. This is clear from the frequency domain picture shown in Fig. \ref{fft}. This is expected since as the system size increases, it approaches the classical limit, thereby suppressing quantum behaviour.


	\bibliography{references}

\begin{thebibliography}{55}%
\makeatletter
\providecommand \@ifxundefined [1]{%
 \@ifx{#1\undefined}
}%
\providecommand \@ifnum [1]{%
 \ifnum #1\expandafter \@firstoftwo
 \else \expandafter \@secondoftwo
 \fi
}%
\providecommand \@ifx [1]{%
 \ifx #1\expandafter \@firstoftwo
 \else \expandafter \@secondoftwo
 \fi
}%
\providecommand \natexlab [1]{#1}%
\providecommand \enquote  [1]{``#1''}%
\providecommand \bibnamefont  [1]{#1}%
\providecommand \bibfnamefont [1]{#1}%
\providecommand \citenamefont [1]{#1}%
\providecommand \href@noop [0]{\@secondoftwo}%
\providecommand \href [0]{\begingroup \@sanitize@url \@href}%
\providecommand \@href[1]{\@@startlink{#1}\@@href}%
\providecommand \@@href[1]{\endgroup#1\@@endlink}%
\providecommand \@sanitize@url [0]{\catcode `\\12\catcode `\$12\catcode
  `\&12\catcode `\#12\catcode `\^12\catcode `\_12\catcode `\%12\relax}%
\providecommand \@@startlink[1]{}%
\providecommand \@@endlink[0]{}%
\providecommand \url  [0]{\begingroup\@sanitize@url \@url }%
\providecommand \@url [1]{\endgroup\@href {#1}{\urlprefix }}%
\providecommand \urlprefix  [0]{URL }%
\providecommand \Eprint [0]{\href }%
\providecommand \doibase [0]{https://doi.org/}%
\providecommand \selectlanguage [0]{\@gobble}%
\providecommand \bibinfo  [0]{\@secondoftwo}%
\providecommand \bibfield  [0]{\@secondoftwo}%
\providecommand \translation [1]{[#1]}%
\providecommand \BibitemOpen [0]{}%
\providecommand \bibitemStop [0]{}%
\providecommand \bibitemNoStop [0]{.\EOS\space}%
\providecommand \EOS [0]{\spacefactor3000\relax}%
\providecommand \BibitemShut  [1]{\csname bibitem#1\endcsname}%
\let\auto@bib@innerbib\@empty
\bibitem [{\citenamefont {Griffiths}\ and\ \citenamefont
  {Schroeter}(2018)}]{griffiths2018introduction}%
  \BibitemOpen
  \bibfield  {author} {\bibinfo {author} {\bibfnamefont {D.~J.}\ \bibnamefont
  {Griffiths}}\ and\ \bibinfo {author} {\bibfnamefont {D.~F.}\ \bibnamefont
  {Schroeter}},\ }\href@noop {} {\emph {\bibinfo {title} {Introduction to
  quantum mechanics}}}\ (\bibinfo  {publisher} {Cambridge university press},\
  \bibinfo {year} {2018})\BibitemShut {NoStop}%
\bibitem [{\citenamefont {Josephson}(1962)}]{Jospehson1962}%
  \BibitemOpen
  \bibfield  {author} {\bibinfo {author} {\bibfnamefont {B.}~\bibnamefont
  {Josephson}},\ }\bibfield  {title} {\bibinfo {title} {Possible new effects in
  superconductive tunnelling},\ }\href
  {https://doi.org/https://doi.org/10.1016/0031-9163(62)91369-0} {\bibfield
  {journal} {\bibinfo  {journal} {Physics Letters}\ }\textbf {\bibinfo {volume}
  {1}},\ \bibinfo {pages} {251} (\bibinfo {year} {1962})}\BibitemShut {NoStop}%
\bibitem [{\citenamefont {Shapiro}(1963)}]{Shapiro1963}%
  \BibitemOpen
  \bibfield  {author} {\bibinfo {author} {\bibfnamefont {S.}~\bibnamefont
  {Shapiro}},\ }\bibfield  {title} {\bibinfo {title} {Josephson currents in
  superconducting tunneling: The effect of microwaves and other observations},\
  }\href {https://doi.org/10.1103/PhysRevLett.11.80} {\bibfield  {journal}
  {\bibinfo  {journal} {Phys. Rev. Lett.}\ }\textbf {\bibinfo {volume} {11}},\
  \bibinfo {pages} {80} (\bibinfo {year} {1963})}\BibitemShut {NoStop}%
\bibitem [{\citenamefont {Tinkham}(2004)}]{tinkham2004introduction}%
  \BibitemOpen
  \bibfield  {author} {\bibinfo {author} {\bibfnamefont {M.}~\bibnamefont
  {Tinkham}},\ }\href@noop {} {\emph {\bibinfo {title} {Introduction to
  superconductivity}}}\ (\bibinfo  {publisher} {Courier Corporation},\ \bibinfo
  {year} {2004})\BibitemShut {NoStop}%
\bibitem [{\citenamefont {Bedrossian}\ \emph {et~al.}(1989)\citenamefont
  {Bedrossian}, \citenamefont {Chen}, \citenamefont {Mortensen},\ and\
  \citenamefont {Golovchenko}}]{bedrossian1989demonstration}%
  \BibitemOpen
  \bibfield  {author} {\bibinfo {author} {\bibfnamefont {P.}~\bibnamefont
  {Bedrossian}}, \bibinfo {author} {\bibfnamefont {D.}~\bibnamefont {Chen}},
  \bibinfo {author} {\bibfnamefont {K.}~\bibnamefont {Mortensen}},\ and\
  \bibinfo {author} {\bibfnamefont {J.}~\bibnamefont {Golovchenko}},\
  }\bibfield  {title} {\bibinfo {title} {Demonstration of the tunnel-diode
  effect on an atomic scale},\ }\href {https://doi.org/10.1038/342258a0}
  {\bibfield  {journal} {\bibinfo  {journal} {Nature}\ }\textbf {\bibinfo
  {volume} {342}},\ \bibinfo {pages} {258} (\bibinfo {year}
  {1989})}\BibitemShut {NoStop}%
\bibitem [{\citenamefont {Mevel}\ \emph {et~al.}(1993)\citenamefont {Mevel},
  \citenamefont {Breger}, \citenamefont {Trainham}, \citenamefont {Petite},
  \citenamefont {Agostini}, \citenamefont {Migus}, \citenamefont {Chambaret},\
  and\ \citenamefont {Antonetti}}]{Mevel1993tunnelionization}%
  \BibitemOpen
  \bibfield  {author} {\bibinfo {author} {\bibfnamefont {E.}~\bibnamefont
  {Mevel}}, \bibinfo {author} {\bibfnamefont {P.}~\bibnamefont {Breger}},
  \bibinfo {author} {\bibfnamefont {R.}~\bibnamefont {Trainham}}, \bibinfo
  {author} {\bibfnamefont {G.}~\bibnamefont {Petite}}, \bibinfo {author}
  {\bibfnamefont {P.}~\bibnamefont {Agostini}}, \bibinfo {author}
  {\bibfnamefont {A.}~\bibnamefont {Migus}}, \bibinfo {author} {\bibfnamefont
  {J.-P.}\ \bibnamefont {Chambaret}},\ and\ \bibinfo {author} {\bibfnamefont
  {A.}~\bibnamefont {Antonetti}},\ }\bibfield  {title} {\bibinfo {title} {Atoms
  in strong optical fields: Evolution from multiphoton to tunnel ionization},\
  }\href {https://doi.org/10.1103/PhysRevLett.70.406} {\bibfield  {journal}
  {\bibinfo  {journal} {Phys. Rev. Lett.}\ }\textbf {\bibinfo {volume} {70}},\
  \bibinfo {pages} {406} (\bibinfo {year} {1993})}\BibitemShut {NoStop}%
\bibitem [{\citenamefont {Ankerhold}(2007)}]{ankerhold2007quantum}%
  \BibitemOpen
  \bibfield  {author} {\bibinfo {author} {\bibfnamefont {J.}~\bibnamefont
  {Ankerhold}},\ }\href {https://doi.org/https://doi.org/10.1007/3-540-68076-4}
  {\emph {\bibinfo {title} {Quantum tunneling in complex systems: the
  semiclassical approach}}},\ Vol.\ \bibinfo {volume} {224}\ (\bibinfo
  {publisher} {Springer Berlin, Heidelberg},\ \bibinfo {year}
  {2007})\BibitemShut {NoStop}%
\bibitem [{\citenamefont {Chen}(2021)}]{chen2021introduction}%
  \BibitemOpen
  \bibfield  {author} {\bibinfo {author} {\bibfnamefont {J.}~\bibnamefont
  {Chen}},\ }\href@noop {} {\emph {\bibinfo {title} {Introduction to Scanning
  Tunneling Microscopy Third Edition}}},\ Vol.~\bibinfo {volume} {69}\
  (\bibinfo  {publisher} {Oxford University Press, USA},\ \bibinfo {year}
  {2021})\BibitemShut {NoStop}%
\bibitem [{\citenamefont {Tomsovic}\ and\ \citenamefont
  {Ullmo}(1994)}]{tomsovicullmotunneling}%
  \BibitemOpen
  \bibfield  {author} {\bibinfo {author} {\bibfnamefont {S.}~\bibnamefont
  {Tomsovic}}\ and\ \bibinfo {author} {\bibfnamefont {D.}~\bibnamefont
  {Ullmo}},\ }\bibfield  {title} {\bibinfo {title} {Chaos-assisted tunneling},\
  }\href {https://doi.org/10.1103/PhysRevE.50.145} {\bibfield  {journal}
  {\bibinfo  {journal} {Phys. Rev. E}\ }\textbf {\bibinfo {volume} {50}},\
  \bibinfo {pages} {145} (\bibinfo {year} {1994})}\BibitemShut {NoStop}%
\bibitem [{\citenamefont {Keshavamurthy}\ and\ \citenamefont
  {Schlagheck}(2011)}]{keshavamurthy2011dynamical}%
  \BibitemOpen
  \bibfield  {author} {\bibinfo {author} {\bibfnamefont {S.}~\bibnamefont
  {Keshavamurthy}}\ and\ \bibinfo {author} {\bibfnamefont {P.}~\bibnamefont
  {Schlagheck}},\ }\href@noop {} {\emph {\bibinfo {title} {Dynamical tunneling:
  theory and experiment}}}\ (\bibinfo  {publisher} {CRC Press},\ \bibinfo
  {year} {2011})\BibitemShut {NoStop}%
\bibitem [{\citenamefont {Davis}\ and\ \citenamefont
  {Heller}(1981)}]{davis1981quantum}%
  \BibitemOpen
  \bibfield  {author} {\bibinfo {author} {\bibfnamefont {M.~J.}\ \bibnamefont
  {Davis}}\ and\ \bibinfo {author} {\bibfnamefont {E.~J.}\ \bibnamefont
  {Heller}},\ }\bibfield  {title} {\bibinfo {title} {Quantum dynamical
  tunneling in bound states},\ }\href@noop {} {\bibfield  {journal} {\bibinfo
  {journal} {The Journal of Chemical Physics}\ }\textbf {\bibinfo {volume}
  {75}},\ \bibinfo {pages} {246} (\bibinfo {year} {1981})}\BibitemShut
  {NoStop}%
\bibitem [{\citenamefont {Heller}\ and\ \citenamefont
  {Davis}(1981)}]{heller1981quantum}%
  \BibitemOpen
  \bibfield  {author} {\bibinfo {author} {\bibfnamefont {E.~J.}\ \bibnamefont
  {Heller}}\ and\ \bibinfo {author} {\bibfnamefont {M.~J.}\ \bibnamefont
  {Davis}},\ }\bibfield  {title} {\bibinfo {title} {Quantum dynamical tunneling
  in large molecules. a plausible conjecture},\ }\href@noop {} {\bibfield
  {journal} {\bibinfo  {journal} {The Journal of Physical Chemistry}\ }\textbf
  {\bibinfo {volume} {85}},\ \bibinfo {pages} {307} (\bibinfo {year}
  {1981})}\BibitemShut {NoStop}%
\bibitem [{\citenamefont {Peres}(1991)}]{peres}%
  \BibitemOpen
  \bibfield  {author} {\bibinfo {author} {\bibfnamefont {A.}~\bibnamefont
  {Peres}},\ }\bibfield  {title} {\bibinfo {title} {Dynamical quasidegeneracies
  and quantum tunneling},\ }\href {https://doi.org/10.1103/PhysRevLett.67.158}
  {\bibfield  {journal} {\bibinfo  {journal} {Phys. Rev. Lett.}\ }\textbf
  {\bibinfo {volume} {67}},\ \bibinfo {pages} {158} (\bibinfo {year}
  {1991})}\BibitemShut {NoStop}%
\bibitem [{\citenamefont {Tomsovic}(2001)}]{tomsovic2001tunneling}%
  \BibitemOpen
  \bibfield  {author} {\bibinfo {author} {\bibfnamefont {S.}~\bibnamefont
  {Tomsovic}},\ }\bibfield  {title} {\bibinfo {title} {Tunneling and chaos},\
  }\href {https://doi.org/https://doi.org/10.1238/Physica.Topical.090a00162}
  {\bibfield  {journal} {\bibinfo  {journal} {Physica Scripta}\ }\textbf
  {\bibinfo {volume} {2001}},\ \bibinfo {pages} {162} (\bibinfo {year}
  {2001})}\BibitemShut {NoStop}%
\bibitem [{\citenamefont {Brodier}\ \emph {et~al.}(2001)\citenamefont
  {Brodier}, \citenamefont {Schlagheck},\ and\ \citenamefont
  {Ullmo}}]{brodierRAT20001}%
  \BibitemOpen
  \bibfield  {author} {\bibinfo {author} {\bibfnamefont {O.}~\bibnamefont
  {Brodier}}, \bibinfo {author} {\bibfnamefont {P.}~\bibnamefont
  {Schlagheck}},\ and\ \bibinfo {author} {\bibfnamefont {D.}~\bibnamefont
  {Ullmo}},\ }\bibfield  {title} {\bibinfo {title} {Resonance-assisted
  tunneling in near-integrable systems},\ }\href
  {https://doi.org/10.1103/PhysRevLett.87.064101} {\bibfield  {journal}
  {\bibinfo  {journal} {Phys. Rev. Lett.}\ }\textbf {\bibinfo {volume} {87}},\
  \bibinfo {pages} {064101} (\bibinfo {year} {2001})}\BibitemShut {NoStop}%
\bibitem [{\citenamefont {B\"acker}\ \emph {et~al.}(2008)\citenamefont
  {B\"acker}, \citenamefont {Ketzmerick}, \citenamefont {L\"ock}, \citenamefont
  {Robnik}, \citenamefont {Vidmar}, \citenamefont {H\"ohmann}, \citenamefont
  {Kuhl},\ and\ \citenamefont {St\"ockmann}}]{backermushroombilliards}%
  \BibitemOpen
  \bibfield  {author} {\bibinfo {author} {\bibfnamefont {A.}~\bibnamefont
  {B\"acker}}, \bibinfo {author} {\bibfnamefont {R.}~\bibnamefont
  {Ketzmerick}}, \bibinfo {author} {\bibfnamefont {S.}~\bibnamefont {L\"ock}},
  \bibinfo {author} {\bibfnamefont {M.}~\bibnamefont {Robnik}}, \bibinfo
  {author} {\bibfnamefont {G.}~\bibnamefont {Vidmar}}, \bibinfo {author}
  {\bibfnamefont {R.}~\bibnamefont {H\"ohmann}}, \bibinfo {author}
  {\bibfnamefont {U.}~\bibnamefont {Kuhl}},\ and\ \bibinfo {author}
  {\bibfnamefont {H.-J.}\ \bibnamefont {St\"ockmann}},\ }\bibfield  {title}
  {\bibinfo {title} {Dynamical tunneling in mushroom billiards},\ }\href
  {https://doi.org/10.1103/PhysRevLett.100.174103} {\bibfield  {journal}
  {\bibinfo  {journal} {Phys. Rev. Lett.}\ }\textbf {\bibinfo {volume} {100}},\
  \bibinfo {pages} {174103} (\bibinfo {year} {2008})}\BibitemShut {NoStop}%
\bibitem [{\citenamefont {Brodier}\ \emph {et~al.}(2002)\citenamefont
  {Brodier}, \citenamefont {Schlagheck},\ and\ \citenamefont
  {Ullmo}}]{brodier2002resonance}%
  \BibitemOpen
  \bibfield  {author} {\bibinfo {author} {\bibfnamefont {O.}~\bibnamefont
  {Brodier}}, \bibinfo {author} {\bibfnamefont {P.}~\bibnamefont
  {Schlagheck}},\ and\ \bibinfo {author} {\bibfnamefont {D.}~\bibnamefont
  {Ullmo}},\ }\bibfield  {title} {\bibinfo {title} {Resonance-assisted
  tunneling},\ }\href@noop {} {\bibfield  {journal} {\bibinfo  {journal}
  {Annals of Physics}\ }\textbf {\bibinfo {volume} {300}},\ \bibinfo {pages}
  {88} (\bibinfo {year} {2002})}\BibitemShut {NoStop}%
\bibitem [{\citenamefont {L\"ock}\ \emph {et~al.}(2010)\citenamefont {L\"ock},
  \citenamefont {B\"acker}, \citenamefont {Ketzmerick},\ and\ \citenamefont
  {Schlagheck}}]{lockregulartochaotic}%
  \BibitemOpen
  \bibfield  {author} {\bibinfo {author} {\bibfnamefont {S.}~\bibnamefont
  {L\"ock}}, \bibinfo {author} {\bibfnamefont {A.}~\bibnamefont {B\"acker}},
  \bibinfo {author} {\bibfnamefont {R.}~\bibnamefont {Ketzmerick}},\ and\
  \bibinfo {author} {\bibfnamefont {P.}~\bibnamefont {Schlagheck}},\ }\bibfield
   {title} {\bibinfo {title} {Regular-to-chaotic tunneling rates: From the
  quantum to the semiclassical regime},\ }\href
  {https://doi.org/10.1103/PhysRevLett.104.114101} {\bibfield  {journal}
  {\bibinfo  {journal} {Phys. Rev. Lett.}\ }\textbf {\bibinfo {volume} {104}},\
  \bibinfo {pages} {114101} (\bibinfo {year} {2010})}\BibitemShut {NoStop}%
\bibitem [{\citenamefont {Eltschka}\ and\ \citenamefont
  {Schlagheck}(2005)}]{Eltschka2005}%
  \BibitemOpen
  \bibfield  {author} {\bibinfo {author} {\bibfnamefont {C.}~\bibnamefont
  {Eltschka}}\ and\ \bibinfo {author} {\bibfnamefont {P.}~\bibnamefont
  {Schlagheck}},\ }\bibfield  {title} {\bibinfo {title} {Resonance- and
  chaos-assisted tunneling in mixed regular-chaotic systems},\ }\href
  {https://doi.org/10.1103/PhysRevLett.94.014101} {\bibfield  {journal}
  {\bibinfo  {journal} {Phys. Rev. Lett.}\ }\textbf {\bibinfo {volume} {94}},\
  \bibinfo {pages} {014101} (\bibinfo {year} {2005})}\BibitemShut {NoStop}%
\bibitem [{\citenamefont {Gehler}\ \emph {et~al.}(2015)\citenamefont {Gehler},
  \citenamefont {L\"ock}, \citenamefont {Shinohara}, \citenamefont {B\"acker},
  \citenamefont {Ketzmerick}, \citenamefont {Kuhl},\ and\ \citenamefont
  {St\"ockmann}}]{gehlarexperimental}%
  \BibitemOpen
  \bibfield  {author} {\bibinfo {author} {\bibfnamefont {S.}~\bibnamefont
  {Gehler}}, \bibinfo {author} {\bibfnamefont {S.}~\bibnamefont {L\"ock}},
  \bibinfo {author} {\bibfnamefont {S.}~\bibnamefont {Shinohara}}, \bibinfo
  {author} {\bibfnamefont {A.}~\bibnamefont {B\"acker}}, \bibinfo {author}
  {\bibfnamefont {R.}~\bibnamefont {Ketzmerick}}, \bibinfo {author}
  {\bibfnamefont {U.}~\bibnamefont {Kuhl}},\ and\ \bibinfo {author}
  {\bibfnamefont {H.-J.}\ \bibnamefont {St\"ockmann}},\ }\bibfield  {title}
  {\bibinfo {title} {Experimental observation of resonance-assisted
  tunneling},\ }\href {https://doi.org/10.1103/PhysRevLett.115.104101}
  {\bibfield  {journal} {\bibinfo  {journal} {Phys. Rev. Lett.}\ }\textbf
  {\bibinfo {volume} {115}},\ \bibinfo {pages} {104101} (\bibinfo {year}
  {2015})}\BibitemShut {NoStop}%
\bibitem [{\citenamefont {Fritzsch}\ \emph {et~al.}(2019)\citenamefont
  {Fritzsch}, \citenamefont {Ketzmerick},\ and\ \citenamefont
  {B\"acker}}]{fritzsch}%
  \BibitemOpen
  \bibfield  {author} {\bibinfo {author} {\bibfnamefont {F.}~\bibnamefont
  {Fritzsch}}, \bibinfo {author} {\bibfnamefont {R.}~\bibnamefont
  {Ketzmerick}},\ and\ \bibinfo {author} {\bibfnamefont {A.}~\bibnamefont
  {B\"acker}},\ }\bibfield  {title} {\bibinfo {title} {Resonance-assisted
  tunneling in deformed optical microdisks with a mixed phase space},\ }\href
  {https://doi.org/10.1103/PhysRevE.100.042219} {\bibfield  {journal} {\bibinfo
   {journal} {Phys. Rev. E}\ }\textbf {\bibinfo {volume} {100}},\ \bibinfo
  {pages} {042219} (\bibinfo {year} {2019})}\BibitemShut {NoStop}%
\bibitem [{\citenamefont {Dembowski}\ \emph {et~al.}(2000)\citenamefont
  {Dembowski}, \citenamefont {Gr\"af}, \citenamefont {Heine}, \citenamefont
  {Hofferbert}, \citenamefont {Rehfeld},\ and\ \citenamefont
  {Richter}}]{Dembowski2000}%
  \BibitemOpen
  \bibfield  {author} {\bibinfo {author} {\bibfnamefont {C.}~\bibnamefont
  {Dembowski}}, \bibinfo {author} {\bibfnamefont {H.-D.}\ \bibnamefont
  {Gr\"af}}, \bibinfo {author} {\bibfnamefont {A.}~\bibnamefont {Heine}},
  \bibinfo {author} {\bibfnamefont {R.}~\bibnamefont {Hofferbert}}, \bibinfo
  {author} {\bibfnamefont {H.}~\bibnamefont {Rehfeld}},\ and\ \bibinfo {author}
  {\bibfnamefont {A.}~\bibnamefont {Richter}},\ }\bibfield  {title} {\bibinfo
  {title} {First experimental evidence for chaos-assisted tunneling in a
  microwave annular billiard},\ }\href
  {https://doi.org/10.1103/PhysRevLett.84.867} {\bibfield  {journal} {\bibinfo
  {journal} {Phys. Rev. Lett.}\ }\textbf {\bibinfo {volume} {84}},\ \bibinfo
  {pages} {867} (\bibinfo {year} {2000})}\BibitemShut {NoStop}%
\bibitem [{\citenamefont {Steck}\ \emph {et~al.}(2001)\citenamefont {Steck},
  \citenamefont {Oskay},\ and\ \citenamefont {Raizen}}]{raizen}%
  \BibitemOpen
  \bibfield  {author} {\bibinfo {author} {\bibfnamefont {D.~A.}\ \bibnamefont
  {Steck}}, \bibinfo {author} {\bibfnamefont {W.~H.}\ \bibnamefont {Oskay}},\
  and\ \bibinfo {author} {\bibfnamefont {M.~G.}\ \bibnamefont {Raizen}},\
  }\bibfield  {title} {\bibinfo {title} {Observation of chaos-assisted
  tunneling between islands of stability},\ }\href
  {https://doi.org/10.1126/science.1061569} {\bibfield  {journal} {\bibinfo
  {journal} {Science}\ }\textbf {\bibinfo {volume} {293}},\ \bibinfo {pages}
  {274} (\bibinfo {year} {2001})}\BibitemShut {NoStop}%
\bibitem [{\citenamefont {Steck}\ \emph {et~al.}(2002)\citenamefont {Steck},
  \citenamefont {Oskay},\ and\ \citenamefont {Raizen}}]{Steck2001fluctuations}%
  \BibitemOpen
  \bibfield  {author} {\bibinfo {author} {\bibfnamefont {D.~A.}\ \bibnamefont
  {Steck}}, \bibinfo {author} {\bibfnamefont {W.~H.}\ \bibnamefont {Oskay}},\
  and\ \bibinfo {author} {\bibfnamefont {M.~G.}\ \bibnamefont {Raizen}},\
  }\bibfield  {title} {\bibinfo {title} {Fluctuations and decoherence in
  chaos-assisted tunneling},\ }\href
  {https://doi.org/10.1103/PhysRevLett.88.120406} {\bibfield  {journal}
  {\bibinfo  {journal} {Phys. Rev. Lett.}\ }\textbf {\bibinfo {volume} {88}},\
  \bibinfo {pages} {120406} (\bibinfo {year} {2002})}\BibitemShut {NoStop}%
\bibitem [{\citenamefont {Hensinger}\ \emph {et~al.}(2001)\citenamefont
  {Hensinger}, \citenamefont {H{\"a}ffner}, \citenamefont {Browaeys},
  \citenamefont {Heckenberg}, \citenamefont {Helmerson}, \citenamefont
  {McKenzie}, \citenamefont {Milburn}, \citenamefont {Phillips}, \citenamefont
  {Rolston}, \citenamefont {Rubinsztein-Dunlop} \emph
  {et~al.}}]{hensinger2001dynamical}%
  \BibitemOpen
  \bibfield  {author} {\bibinfo {author} {\bibfnamefont {W.~K.}\ \bibnamefont
  {Hensinger}}, \bibinfo {author} {\bibfnamefont {H.}~\bibnamefont
  {H{\"a}ffner}}, \bibinfo {author} {\bibfnamefont {A.}~\bibnamefont
  {Browaeys}}, \bibinfo {author} {\bibfnamefont {N.~R.}\ \bibnamefont
  {Heckenberg}}, \bibinfo {author} {\bibfnamefont {K.}~\bibnamefont
  {Helmerson}}, \bibinfo {author} {\bibfnamefont {C.}~\bibnamefont {McKenzie}},
  \bibinfo {author} {\bibfnamefont {G.~J.}\ \bibnamefont {Milburn}}, \bibinfo
  {author} {\bibfnamefont {W.~D.}\ \bibnamefont {Phillips}}, \bibinfo {author}
  {\bibfnamefont {S.~L.}\ \bibnamefont {Rolston}}, \bibinfo {author}
  {\bibfnamefont {H.}~\bibnamefont {Rubinsztein-Dunlop}}, \emph {et~al.},\
  }\bibfield  {title} {\bibinfo {title} {Dynamical tunnelling of ultracold
  atoms},\ }\href@noop {} {\bibfield  {journal} {\bibinfo  {journal} {Nature}\
  }\textbf {\bibinfo {volume} {412}},\ \bibinfo {pages} {52} (\bibinfo {year}
  {2001})}\BibitemShut {NoStop}%
\bibitem [{\citenamefont {Mouchet}\ \emph {et~al.}(2001)\citenamefont
  {Mouchet}, \citenamefont {Miniatura}, \citenamefont {Kaiser}, \citenamefont
  {Gr\'emaud},\ and\ \citenamefont {Delande}}]{Mouchet}%
  \BibitemOpen
  \bibfield  {author} {\bibinfo {author} {\bibfnamefont {A.}~\bibnamefont
  {Mouchet}}, \bibinfo {author} {\bibfnamefont {C.}~\bibnamefont {Miniatura}},
  \bibinfo {author} {\bibfnamefont {R.}~\bibnamefont {Kaiser}}, \bibinfo
  {author} {\bibfnamefont {B.}~\bibnamefont {Gr\'emaud}},\ and\ \bibinfo
  {author} {\bibfnamefont {D.}~\bibnamefont {Delande}},\ }\bibfield  {title}
  {\bibinfo {title} {Chaos-assisted tunneling with cold atoms},\ }\href
  {https://doi.org/10.1103/PhysRevE.64.016221} {\bibfield  {journal} {\bibinfo
  {journal} {Phys. Rev. E}\ }\textbf {\bibinfo {volume} {64}},\ \bibinfo
  {pages} {016221} (\bibinfo {year} {2001})}\BibitemShut {NoStop}%
\bibitem [{\citenamefont {Hofferbert}\ \emph {et~al.}(2005)\citenamefont
  {Hofferbert}, \citenamefont {Alt}, \citenamefont {Dembowski}, \citenamefont
  {Gr\"af}, \citenamefont {Harney}, \citenamefont {Heine}, \citenamefont
  {Rehfeld},\ and\ \citenamefont {Richter}}]{Hofferbert2005}%
  \BibitemOpen
  \bibfield  {author} {\bibinfo {author} {\bibfnamefont {R.}~\bibnamefont
  {Hofferbert}}, \bibinfo {author} {\bibfnamefont {H.}~\bibnamefont {Alt}},
  \bibinfo {author} {\bibfnamefont {C.}~\bibnamefont {Dembowski}}, \bibinfo
  {author} {\bibfnamefont {H.-D.}\ \bibnamefont {Gr\"af}}, \bibinfo {author}
  {\bibfnamefont {H.~L.}\ \bibnamefont {Harney}}, \bibinfo {author}
  {\bibfnamefont {A.}~\bibnamefont {Heine}}, \bibinfo {author} {\bibfnamefont
  {H.}~\bibnamefont {Rehfeld}},\ and\ \bibinfo {author} {\bibfnamefont
  {A.}~\bibnamefont {Richter}},\ }\bibfield  {title} {\bibinfo {title}
  {Experimental investigations of chaos-assisted tunneling in a microwave
  annular billiard},\ }\href {https://doi.org/10.1103/PhysRevE.71.046201}
  {\bibfield  {journal} {\bibinfo  {journal} {Phys. Rev. E}\ }\textbf {\bibinfo
  {volume} {71}},\ \bibinfo {pages} {046201} (\bibinfo {year}
  {2005})}\BibitemShut {NoStop}%
\bibitem [{\citenamefont {Vanhaele}\ and\ \citenamefont
  {Schlagheck}(2021)}]{Schlagheck2021}%
  \BibitemOpen
  \bibfield  {author} {\bibinfo {author} {\bibfnamefont {G.}~\bibnamefont
  {Vanhaele}}\ and\ \bibinfo {author} {\bibfnamefont {P.}~\bibnamefont
  {Schlagheck}},\ }\bibfield  {title} {\bibinfo {title} {Noon states with
  ultracold bosonic atoms via resonance- and chaos-assisted tunneling},\ }\href
  {https://doi.org/10.1103/PhysRevA.103.013315} {\bibfield  {journal} {\bibinfo
   {journal} {Phys. Rev. A}\ }\textbf {\bibinfo {volume} {103}},\ \bibinfo
  {pages} {013315} (\bibinfo {year} {2021})}\BibitemShut {NoStop}%
\bibitem [{\citenamefont {Vanhaele}\ \emph {et~al.}(2022)\citenamefont
  {Vanhaele}, \citenamefont {B\"acker}, \citenamefont {Ketzmerick},\ and\
  \citenamefont {Schlagheck}}]{VanhaeleNOON2022}%
  \BibitemOpen
  \bibfield  {author} {\bibinfo {author} {\bibfnamefont {G.}~\bibnamefont
  {Vanhaele}}, \bibinfo {author} {\bibfnamefont {A.}~\bibnamefont {B\"acker}},
  \bibinfo {author} {\bibfnamefont {R.}~\bibnamefont {Ketzmerick}},\ and\
  \bibinfo {author} {\bibfnamefont {P.}~\bibnamefont {Schlagheck}},\ }\bibfield
   {title} {\bibinfo {title} {Creating triple-noon states with ultracold atoms
  via chaos-assisted tunneling},\ }\href
  {https://doi.org/10.1103/PhysRevA.106.L011301} {\bibfield  {journal}
  {\bibinfo  {journal} {Phys. Rev. A}\ }\textbf {\bibinfo {volume} {106}},\
  \bibinfo {pages} {L011301} (\bibinfo {year} {2022})}\BibitemShut {NoStop}%
\bibitem [{\citenamefont {Satpathi}\ \emph {et~al.}(2022)\citenamefont
  {Satpathi}, \citenamefont {Ray},\ and\ \citenamefont {Vardi}}]{UrbashiS2022}%
  \BibitemOpen
  \bibfield  {author} {\bibinfo {author} {\bibfnamefont {U.}~\bibnamefont
  {Satpathi}}, \bibinfo {author} {\bibfnamefont {S.}~\bibnamefont {Ray}},\ and\
  \bibinfo {author} {\bibfnamefont {A.}~\bibnamefont {Vardi}},\ }\bibfield
  {title} {\bibinfo {title} {Chaos-assisted many-body tunnelling},\ }\href
  {https://doi.org/10.1103/PhysRevE.106.L042204} {\bibfield  {journal}
  {\bibinfo  {journal} {Phys. Rev. E}\ }\textbf {\bibinfo {volume} {106}},\
  \bibinfo {pages} {L042204} (\bibinfo {year} {2022})}\BibitemShut {NoStop}%
\bibitem [{\citenamefont {Haake}\ \emph {et~al.}(1987)\citenamefont {Haake},
  \citenamefont {Ku{\'s}},\ and\ \citenamefont {Scharf}}]{haake1987classical}%
  \BibitemOpen
  \bibfield  {author} {\bibinfo {author} {\bibfnamefont {F.}~\bibnamefont
  {Haake}}, \bibinfo {author} {\bibfnamefont {M.}~\bibnamefont {Ku{\'s}}},\
  and\ \bibinfo {author} {\bibfnamefont {R.}~\bibnamefont {Scharf}},\
  }\bibfield  {title} {\bibinfo {title} {Classical and quantum chaos for a
  kicked top},\ }\href@noop {} {\bibfield  {journal} {\bibinfo  {journal}
  {Zeitschrift f{\"u}r Physik B Condensed Matter}\ }\textbf {\bibinfo {volume}
  {65}},\ \bibinfo {pages} {381} (\bibinfo {year} {1987})}\BibitemShut
  {NoStop}%
\bibitem [{\citenamefont {Chaudhury}\ \emph {et~al.}(2009)\citenamefont
  {Chaudhury}, \citenamefont {Smith}, \citenamefont {Anderson}, \citenamefont
  {Ghose},\ and\ \citenamefont {Jessen}}]{chaudhury2009quantum}%
  \BibitemOpen
  \bibfield  {author} {\bibinfo {author} {\bibfnamefont {S.}~\bibnamefont
  {Chaudhury}}, \bibinfo {author} {\bibfnamefont {A.}~\bibnamefont {Smith}},
  \bibinfo {author} {\bibfnamefont {B.}~\bibnamefont {Anderson}}, \bibinfo
  {author} {\bibfnamefont {S.}~\bibnamefont {Ghose}},\ and\ \bibinfo {author}
  {\bibfnamefont {P.~S.}\ \bibnamefont {Jessen}},\ }\bibfield  {title}
  {\bibinfo {title} {Quantum signatures of chaos in a kicked top},\ }\href
  {https://doi.org/10.1038/nature08396} {\bibfield  {journal} {\bibinfo
  {journal} {Nature}\ }\textbf {\bibinfo {volume} {461}},\ \bibinfo {pages}
  {768} (\bibinfo {year} {2009})}\BibitemShut {NoStop}%
\bibitem [{\citenamefont {Sanders}\ and\ \citenamefont
  {Milburn}(1989)}]{sanders1989effect}%
  \BibitemOpen
  \bibfield  {author} {\bibinfo {author} {\bibfnamefont {B.}~\bibnamefont
  {Sanders}}\ and\ \bibinfo {author} {\bibfnamefont {G.}~\bibnamefont
  {Milburn}},\ }\bibfield  {title} {\bibinfo {title} {The effect of measurement
  on the quantum features of a chaotic system},\ }\href@noop {} {\bibfield
  {journal} {\bibinfo  {journal} {Zeitschrift f{\"u}r Physik B Condensed
  Matter}\ }\textbf {\bibinfo {volume} {77}},\ \bibinfo {pages} {497} (\bibinfo
  {year} {1989})}\BibitemShut {NoStop}%
\bibitem [{\citenamefont {Dogra}\ \emph {et~al.}(2019)\citenamefont {Dogra},
  \citenamefont {Madhok},\ and\ \citenamefont {Lakshminarayan}}]{Dogra2019}%
  \BibitemOpen
  \bibfield  {author} {\bibinfo {author} {\bibfnamefont {S.}~\bibnamefont
  {Dogra}}, \bibinfo {author} {\bibfnamefont {V.}~\bibnamefont {Madhok}},\ and\
  \bibinfo {author} {\bibfnamefont {A.}~\bibnamefont {Lakshminarayan}},\
  }\bibfield  {title} {\bibinfo {title} {Quantum signatures of chaos,
  thermalization, and tunneling in the exactly solvable few-body kicked top},\
  }\href {https://doi.org/10.1103/PhysRevE.99.062217} {\bibfield  {journal}
  {\bibinfo  {journal} {Phys. Rev. E}\ }\textbf {\bibinfo {volume} {99}},\
  \bibinfo {pages} {062217} (\bibinfo {year} {2019})}\BibitemShut {NoStop}%
\bibitem [{\citenamefont {Cory}\ \emph {et~al.}(2000)\citenamefont {Cory},
  \citenamefont {Laflamme}, \citenamefont {Knill}, \citenamefont {Viola},
  \citenamefont {Havel}, \citenamefont {Boulant}, \citenamefont {Boutis},
  \citenamefont {Fortunato}, \citenamefont {Lloyd}, \citenamefont {Martinez}
  \emph {et~al.}}]{cory2000nmr}%
  \BibitemOpen
  \bibfield  {author} {\bibinfo {author} {\bibfnamefont {D.~G.}\ \bibnamefont
  {Cory}}, \bibinfo {author} {\bibfnamefont {R.}~\bibnamefont {Laflamme}},
  \bibinfo {author} {\bibfnamefont {E.}~\bibnamefont {Knill}}, \bibinfo
  {author} {\bibfnamefont {L.}~\bibnamefont {Viola}}, \bibinfo {author}
  {\bibfnamefont {T.}~\bibnamefont {Havel}}, \bibinfo {author} {\bibfnamefont
  {N.}~\bibnamefont {Boulant}}, \bibinfo {author} {\bibfnamefont
  {G.}~\bibnamefont {Boutis}}, \bibinfo {author} {\bibfnamefont
  {E.}~\bibnamefont {Fortunato}}, \bibinfo {author} {\bibfnamefont
  {S.}~\bibnamefont {Lloyd}}, \bibinfo {author} {\bibfnamefont
  {R.}~\bibnamefont {Martinez}}, \emph {et~al.},\ }\bibfield  {title} {\bibinfo
  {title} {Nmr based quantum information processing: Achievements and
  prospects},\ }\href
  {https://doi.org/https://doi.org/10.1002/1521-3978(200009)48:9/11<875::AID-PROP875>3.0.CO;2-V}
  {\bibfield  {journal} {\bibinfo  {journal} {Fortschritte der Physik: Progress
  of Physics}\ }\textbf {\bibinfo {volume} {48}},\ \bibinfo {pages} {875}
  (\bibinfo {year} {2000})}\BibitemShut {NoStop}%
\bibitem [{\citenamefont {Suter}\ and\ \citenamefont
  {Mahesh}(2008)}]{suter2008spins}%
  \BibitemOpen
  \bibfield  {author} {\bibinfo {author} {\bibfnamefont {D.}~\bibnamefont
  {Suter}}\ and\ \bibinfo {author} {\bibfnamefont {T.}~\bibnamefont {Mahesh}},\
  }\bibfield  {title} {\bibinfo {title} {Spins as qubits: quantum information
  processing by nuclear magnetic resonance},\ }\href
  {https://doi.org/10.1063/1.2838166} {\bibfield  {journal} {\bibinfo
  {journal} {The Journal of chemical physics}\ }\textbf {\bibinfo {volume}
  {128}},\ \bibinfo {pages} {052206} (\bibinfo {year} {2008})},\ \Eprint
  {https://arxiv.org/abs/https://doi.org/10.1063/1.2838166}
  {https://doi.org/10.1063/1.2838166} \BibitemShut {NoStop}%
\bibitem [{\citenamefont {Araujo-Ferreira}\ \emph {et~al.}(2013)\citenamefont
  {Araujo-Ferreira}, \citenamefont {Auccaise}, \citenamefont {Sarthour},
  \citenamefont {Oliveira}, \citenamefont {Bonagamba},\ and\ \citenamefont
  {Roditi}}]{araujo2013classical}%
  \BibitemOpen
  \bibfield  {author} {\bibinfo {author} {\bibfnamefont {A.}~\bibnamefont
  {Araujo-Ferreira}}, \bibinfo {author} {\bibfnamefont {R.}~\bibnamefont
  {Auccaise}}, \bibinfo {author} {\bibfnamefont {R.}~\bibnamefont {Sarthour}},
  \bibinfo {author} {\bibfnamefont {I.}~\bibnamefont {Oliveira}}, \bibinfo
  {author} {\bibfnamefont {T.~J.}\ \bibnamefont {Bonagamba}},\ and\ \bibinfo
  {author} {\bibfnamefont {I.}~\bibnamefont {Roditi}},\ }\bibfield  {title}
  {\bibinfo {title} {Classical bifurcation in a quadrupolar nmr system},\
  }\href {https://doi.org/10.1103/PhysRevA.87.053605} {\bibfield  {journal}
  {\bibinfo  {journal} {Physical Review A}\ }\textbf {\bibinfo {volume} {87}},\
  \bibinfo {pages} {053605} (\bibinfo {year} {2013})}\BibitemShut {NoStop}%
\bibitem [{\citenamefont {Krithika}\ \emph {et~al.}(2019)\citenamefont
  {Krithika}, \citenamefont {Anjusha}, \citenamefont {Bhosale},\ and\
  \citenamefont {Mahesh}}]{krithika2019}%
  \BibitemOpen
  \bibfield  {author} {\bibinfo {author} {\bibfnamefont {V.~R.}\ \bibnamefont
  {Krithika}}, \bibinfo {author} {\bibfnamefont {V.~S.}\ \bibnamefont
  {Anjusha}}, \bibinfo {author} {\bibfnamefont {U.~T.}\ \bibnamefont
  {Bhosale}},\ and\ \bibinfo {author} {\bibfnamefont {T.~S.}\ \bibnamefont
  {Mahesh}},\ }\bibfield  {title} {\bibinfo {title} {Nmr studies of quantum
  chaos in a two-qubit kicked top},\ }\href
  {https://doi.org/10.1103/PhysRevE.99.032219} {\bibfield  {journal} {\bibinfo
  {journal} {Phys. Rev. E}\ }\textbf {\bibinfo {volume} {99}},\ \bibinfo
  {pages} {032219} (\bibinfo {year} {2019})}\BibitemShut {NoStop}%
\bibitem [{\citenamefont {Krithika}\ \emph {et~al.}(2022)\citenamefont
  {Krithika}, \citenamefont {Solanki}, \citenamefont {Vinjanampathy},\ and\
  \citenamefont {Mahesh}}]{krithika2022observation}%
  \BibitemOpen
  \bibfield  {author} {\bibinfo {author} {\bibfnamefont {V.}~\bibnamefont
  {Krithika}}, \bibinfo {author} {\bibfnamefont {P.}~\bibnamefont {Solanki}},
  \bibinfo {author} {\bibfnamefont {S.}~\bibnamefont {Vinjanampathy}},\ and\
  \bibinfo {author} {\bibfnamefont {T.}~\bibnamefont {Mahesh}},\ }\bibfield
  {title} {\bibinfo {title} {Observation of quantum phase synchronization in a
  nuclear-spin system},\ }\href {https://doi.org/10.1103/PhysRevA.105.062206}
  {\bibfield  {journal} {\bibinfo  {journal} {Physical Review A}\ }\textbf
  {\bibinfo {volume} {105}},\ \bibinfo {pages} {062206} (\bibinfo {year}
  {2022})}\BibitemShut {NoStop}%
\bibitem [{\citenamefont {Peng}\ \emph {et~al.}(2005)\citenamefont {Peng},
  \citenamefont {Du},\ and\ \citenamefont {Suter}}]{Suter2005QPT}%
  \BibitemOpen
  \bibfield  {author} {\bibinfo {author} {\bibfnamefont {X.}~\bibnamefont
  {Peng}}, \bibinfo {author} {\bibfnamefont {J.}~\bibnamefont {Du}},\ and\
  \bibinfo {author} {\bibfnamefont {D.}~\bibnamefont {Suter}},\ }\bibfield
  {title} {\bibinfo {title} {Quantum phase transition of ground-state
  entanglement in a heisenberg spin chain simulated in an nmr quantum
  computer},\ }\href {https://doi.org/10.1103/PhysRevA.71.012307} {\bibfield
  {journal} {\bibinfo  {journal} {Phys. Rev. A}\ }\textbf {\bibinfo {volume}
  {71}},\ \bibinfo {pages} {012307} (\bibinfo {year} {2005})}\BibitemShut
  {NoStop}%
\bibitem [{\citenamefont {Fortes}\ \emph {et~al.}(2020)\citenamefont {Fortes},
  \citenamefont {García-Mata}, \citenamefont {Jalabert},\ and\ \citenamefont
  {Wisniacki}}]{Fortes2020}%
  \BibitemOpen
  \bibfield  {author} {\bibinfo {author} {\bibfnamefont {E.~M.}\ \bibnamefont
  {Fortes}}, \bibinfo {author} {\bibfnamefont {I.}~\bibnamefont
  {García-Mata}}, \bibinfo {author} {\bibfnamefont {R.~A.}\ \bibnamefont
  {Jalabert}},\ and\ \bibinfo {author} {\bibfnamefont {D.~A.}\ \bibnamefont
  {Wisniacki}},\ }\bibfield  {title} {\bibinfo {title} {Signatures of quantum
  chaos transition in short spin chains},\ }\href
  {https://doi.org/10.1209/0295-5075/130/60001} {\bibfield  {journal} {\bibinfo
   {journal} {Europhysics Letters}\ }\textbf {\bibinfo {volume} {130}},\
  \bibinfo {pages} {60001} (\bibinfo {year} {2020})}\BibitemShut {NoStop}%
\bibitem [{\citenamefont {Nie}\ \emph {et~al.}(2020)\citenamefont {Nie},
  \citenamefont {Wei}, \citenamefont {Chen}, \citenamefont {Zhang},
  \citenamefont {Zhao}, \citenamefont {Qiu}, \citenamefont {Tian},
  \citenamefont {Ji}, \citenamefont {Xin}, \citenamefont {Lu},\ and\
  \citenamefont {Li}}]{Li2020}%
  \BibitemOpen
  \bibfield  {author} {\bibinfo {author} {\bibfnamefont {X.}~\bibnamefont
  {Nie}}, \bibinfo {author} {\bibfnamefont {B.-B.}\ \bibnamefont {Wei}},
  \bibinfo {author} {\bibfnamefont {X.}~\bibnamefont {Chen}}, \bibinfo {author}
  {\bibfnamefont {Z.}~\bibnamefont {Zhang}}, \bibinfo {author} {\bibfnamefont
  {X.}~\bibnamefont {Zhao}}, \bibinfo {author} {\bibfnamefont {C.}~\bibnamefont
  {Qiu}}, \bibinfo {author} {\bibfnamefont {Y.}~\bibnamefont {Tian}}, \bibinfo
  {author} {\bibfnamefont {Y.}~\bibnamefont {Ji}}, \bibinfo {author}
  {\bibfnamefont {T.}~\bibnamefont {Xin}}, \bibinfo {author} {\bibfnamefont
  {D.}~\bibnamefont {Lu}},\ and\ \bibinfo {author} {\bibfnamefont
  {J.}~\bibnamefont {Li}},\ }\bibfield  {title} {\bibinfo {title} {Experimental
  observation of equilibrium and dynamical quantum phase transitions via
  out-of-time-ordered correlators},\ }\href
  {https://doi.org/10.1103/PhysRevLett.124.250601} {\bibfield  {journal}
  {\bibinfo  {journal} {Phys. Rev. Lett.}\ }\textbf {\bibinfo {volume} {124}},\
  \bibinfo {pages} {250601} (\bibinfo {year} {2020})}\BibitemShut {NoStop}%
\bibitem [{\citenamefont {Li}\ \emph {et~al.}(2017)\citenamefont {Li},
  \citenamefont {Fan}, \citenamefont {Wang}, \citenamefont {Ye}, \citenamefont
  {Zeng}, \citenamefont {Zhai}, \citenamefont {Peng},\ and\ \citenamefont
  {Du}}]{DuOTOC}%
  \BibitemOpen
  \bibfield  {author} {\bibinfo {author} {\bibfnamefont {J.}~\bibnamefont
  {Li}}, \bibinfo {author} {\bibfnamefont {R.}~\bibnamefont {Fan}}, \bibinfo
  {author} {\bibfnamefont {H.}~\bibnamefont {Wang}}, \bibinfo {author}
  {\bibfnamefont {B.}~\bibnamefont {Ye}}, \bibinfo {author} {\bibfnamefont
  {B.}~\bibnamefont {Zeng}}, \bibinfo {author} {\bibfnamefont {H.}~\bibnamefont
  {Zhai}}, \bibinfo {author} {\bibfnamefont {X.}~\bibnamefont {Peng}},\ and\
  \bibinfo {author} {\bibfnamefont {J.}~\bibnamefont {Du}},\ }\bibfield
  {title} {\bibinfo {title} {Measuring out-of-time-order correlators on a
  nuclear magnetic resonance quantum simulator},\ }\href
  {https://doi.org/10.1103/PhysRevX.7.031011} {\bibfield  {journal} {\bibinfo
  {journal} {Phys. Rev. X}\ }\textbf {\bibinfo {volume} {7}},\ \bibinfo {pages}
  {031011} (\bibinfo {year} {2017})}\BibitemShut {NoStop}%
\bibitem [{\citenamefont {Bhosale}\ and\ \citenamefont
  {Santhanam}(2017)}]{UdaySignatures}%
  \BibitemOpen
  \bibfield  {author} {\bibinfo {author} {\bibfnamefont {U.~T.}\ \bibnamefont
  {Bhosale}}\ and\ \bibinfo {author} {\bibfnamefont {M.~S.}\ \bibnamefont
  {Santhanam}},\ }\bibfield  {title} {\bibinfo {title} {Signatures of
  bifurcation on quantum correlations: Case of the quantum kicked top},\ }\href
  {https://doi.org/10.1103/PhysRevE.95.012216} {\bibfield  {journal} {\bibinfo
  {journal} {Phys. Rev. E}\ }\textbf {\bibinfo {volume} {95}},\ \bibinfo
  {pages} {012216} (\bibinfo {year} {2017})}\BibitemShut {NoStop}%
\bibitem [{\citenamefont {Levitt}(2013)}]{levitt2013spin}%
  \BibitemOpen
  \bibfield  {author} {\bibinfo {author} {\bibfnamefont {M.~H.}\ \bibnamefont
  {Levitt}},\ }\href@noop {} {\emph {\bibinfo {title} {Spin dynamics: basics of
  nuclear magnetic resonance}}}\ (\bibinfo  {publisher} {John Wiley \& Sons},\
  \bibinfo {year} {2013})\BibitemShut {NoStop}%
\bibitem [{\citenamefont {Cavanagh}\ \emph {et~al.}(1995)\citenamefont
  {Cavanagh}, \citenamefont {Fairbrother}, \citenamefont {III},\ and\
  \citenamefont {Skelton}}]{cavanagh}%
  \BibitemOpen
  \bibfield  {author} {\bibinfo {author} {\bibfnamefont {J.}~\bibnamefont
  {Cavanagh}}, \bibinfo {author} {\bibfnamefont {W.~J.}\ \bibnamefont
  {Fairbrother}}, \bibinfo {author} {\bibfnamefont {A.~G.~P.}\ \bibnamefont
  {III}},\ and\ \bibinfo {author} {\bibfnamefont {N.~J.}\ \bibnamefont
  {Skelton}},\ }\href@noop {} {\emph {\bibinfo {title} {Protein NMR
  spectroscopy: principles and practice}}}\ (\bibinfo  {publisher} {Elsevier},\
  \bibinfo {year} {1995})\BibitemShut {NoStop}%
\bibitem [{\citenamefont {Wang}\ \emph {et~al.}(2004)\citenamefont {Wang},
  \citenamefont {Ghose}, \citenamefont {Sanders},\ and\ \citenamefont
  {Hu}}]{Wang2004}%
  \BibitemOpen
  \bibfield  {author} {\bibinfo {author} {\bibfnamefont {X.}~\bibnamefont
  {Wang}}, \bibinfo {author} {\bibfnamefont {S.}~\bibnamefont {Ghose}},
  \bibinfo {author} {\bibfnamefont {B.~C.}\ \bibnamefont {Sanders}},\ and\
  \bibinfo {author} {\bibfnamefont {B.}~\bibnamefont {Hu}},\ }\bibfield
  {title} {\bibinfo {title} {Entanglement as a signature of quantum chaos},\
  }\href {https://doi.org/10.1103/PhysRevE.70.016217} {\bibfield  {journal}
  {\bibinfo  {journal} {Phys. Rev. E}\ }\textbf {\bibinfo {volume} {70}},\
  \bibinfo {pages} {016217} (\bibinfo {year} {2004})}\BibitemShut {NoStop}%
\bibitem [{\citenamefont {Radcliffe}(1971)}]{Radcliffe1971}%
  \BibitemOpen
  \bibfield  {author} {\bibinfo {author} {\bibfnamefont {J.~M.}\ \bibnamefont
  {Radcliffe}},\ }\bibfield  {title} {\bibinfo {title} {Some properties of
  coherent spin states},\ }\href {https://doi.org/10.1088/0305-4470/4/3/009}
  {\bibfield  {journal} {\bibinfo  {journal} {Journal of Physics A: General
  Physics}\ }\textbf {\bibinfo {volume} {4}},\ \bibinfo {pages} {313} (\bibinfo
  {year} {1971})}\BibitemShut {NoStop}%
\bibitem [{\citenamefont {Klauder}\ and\ \citenamefont
  {Skagerstam}(1985)}]{KlauderCoherent}%
  \BibitemOpen
  \bibfield  {author} {\bibinfo {author} {\bibfnamefont {J.~R.}\ \bibnamefont
  {Klauder}}\ and\ \bibinfo {author} {\bibfnamefont {B.}~\bibnamefont
  {Skagerstam}},\ }\href@noop {} {\emph {\bibinfo {title} {Coherent States:
  Applications in Physics and Mathematical Physics}}}\ (\bibinfo  {publisher}
  {World Scientific},\ \bibinfo {year} {1985})\BibitemShut {NoStop}%
\bibitem [{\citenamefont {Cory}\ \emph {et~al.}(1997)\citenamefont {Cory},
  \citenamefont {Fahmy},\ and\ \citenamefont {Havel}}]{CoryPPS}%
  \BibitemOpen
  \bibfield  {author} {\bibinfo {author} {\bibfnamefont {D.~G.}\ \bibnamefont
  {Cory}}, \bibinfo {author} {\bibfnamefont {A.~F.}\ \bibnamefont {Fahmy}},\
  and\ \bibinfo {author} {\bibfnamefont {T.~F.}\ \bibnamefont {Havel}},\
  }\bibfield  {title} {\bibinfo {title} {Ensemble quantum computing by nmr
  spectroscopy},\ }\href {https://doi.org/10.1073/pnas.94.5.1634} {\bibfield
  {journal} {\bibinfo  {journal} {Proceedings of the National Academy of
  Sciences}\ }\textbf {\bibinfo {volume} {94}},\ \bibinfo {pages} {1634}
  (\bibinfo {year} {1997})}\BibitemShut {NoStop}%
\bibitem [{\citenamefont {Gershenfeld}\ and\ \citenamefont
  {Chuang}(1997)}]{gershenfeld1997bulk}%
  \BibitemOpen
  \bibfield  {author} {\bibinfo {author} {\bibfnamefont {N.~A.}\ \bibnamefont
  {Gershenfeld}}\ and\ \bibinfo {author} {\bibfnamefont {I.~L.}\ \bibnamefont
  {Chuang}},\ }\bibfield  {title} {\bibinfo {title} {Bulk spin-resonance
  quantum computation},\ }\href {https://doi.org/10.1126/science.275.5298.350}
  {\bibfield  {journal} {\bibinfo  {journal} {Science}\ }\textbf {\bibinfo
  {volume} {275}},\ \bibinfo {pages} {350} (\bibinfo {year}
  {1997})}\BibitemShut {NoStop}%
\bibitem [{\citenamefont {Krithika}\ \emph {et~al.}(2021)\citenamefont
  {Krithika}, \citenamefont {Pal}, \citenamefont {Nath},\ and\ \citenamefont
  {Mahesh}}]{krithikaBlockade}%
  \BibitemOpen
  \bibfield  {author} {\bibinfo {author} {\bibfnamefont {V.~R.}\ \bibnamefont
  {Krithika}}, \bibinfo {author} {\bibfnamefont {S.}~\bibnamefont {Pal}},
  \bibinfo {author} {\bibfnamefont {R.}~\bibnamefont {Nath}},\ and\ \bibinfo
  {author} {\bibfnamefont {T.~S.}\ \bibnamefont {Mahesh}},\ }\bibfield  {title}
  {\bibinfo {title} {Observation of interaction induced blockade and local spin
  freezing in a nmr quantum simulator},\ }\href
  {https://doi.org/10.1103/PhysRevResearch.3.033035} {\bibfield  {journal}
  {\bibinfo  {journal} {Phys. Rev. Research}\ }\textbf {\bibinfo {volume}
  {3}},\ \bibinfo {pages} {033035} (\bibinfo {year} {2021})}\BibitemShut
  {NoStop}%
\bibitem [{\citenamefont {Fortunato}\ \emph {et~al.}(2002)\citenamefont
  {Fortunato}, \citenamefont {Pravia}, \citenamefont {Boulant}, \citenamefont
  {Teklemariam}, \citenamefont {Havel},\ and\ \citenamefont
  {Cory}}]{Fortunato}%
  \BibitemOpen
  \bibfield  {author} {\bibinfo {author} {\bibfnamefont {E.~M.}\ \bibnamefont
  {Fortunato}}, \bibinfo {author} {\bibfnamefont {M.~A.}\ \bibnamefont
  {Pravia}}, \bibinfo {author} {\bibfnamefont {N.}~\bibnamefont {Boulant}},
  \bibinfo {author} {\bibfnamefont {G.}~\bibnamefont {Teklemariam}}, \bibinfo
  {author} {\bibfnamefont {T.~F.}\ \bibnamefont {Havel}},\ and\ \bibinfo
  {author} {\bibfnamefont {D.~G.}\ \bibnamefont {Cory}},\ }\bibfield  {title}
  {\bibinfo {title} {Design of strongly modulating pulses to implement precise
  effective hamiltonians for quantum information processing},\ }\href
  {https://doi.org/10.1063/1.1465412} {\bibfield  {journal} {\bibinfo
  {journal} {The Journal of Chemical Physics}\ }\textbf {\bibinfo {volume}
  {116}},\ \bibinfo {pages} {7599} (\bibinfo {year} {2002})},\ \Eprint
  {https://arxiv.org/abs/https://doi.org/10.1063/1.1465412}
  {https://doi.org/10.1063/1.1465412} \BibitemShut {NoStop}%
\bibitem [{\citenamefont {Chen}\ \emph {et~al.}(2015)\citenamefont {Chen},
  \citenamefont {An}, \citenamefont {Luo}, \citenamefont {Sun},\ and\
  \citenamefont {Oh}}]{FloquetEngg1}%
  \BibitemOpen
  \bibfield  {author} {\bibinfo {author} {\bibfnamefont {C.}~\bibnamefont
  {Chen}}, \bibinfo {author} {\bibfnamefont {J.-H.}\ \bibnamefont {An}},
  \bibinfo {author} {\bibfnamefont {H.-G.}\ \bibnamefont {Luo}}, \bibinfo
  {author} {\bibfnamefont {C.~P.}\ \bibnamefont {Sun}},\ and\ \bibinfo {author}
  {\bibfnamefont {C.~H.}\ \bibnamefont {Oh}},\ }\bibfield  {title} {\bibinfo
  {title} {Floquet control of quantum dissipation in spin chains},\ }\href
  {https://doi.org/10.1103/PhysRevA.91.052122} {\bibfield  {journal} {\bibinfo
  {journal} {Phys. Rev. A}\ }\textbf {\bibinfo {volume} {91}},\ \bibinfo
  {pages} {052122} (\bibinfo {year} {2015})}\BibitemShut {NoStop}%
\bibitem [{\citenamefont {Claeys}\ \emph {et~al.}(2019)\citenamefont {Claeys},
  \citenamefont {Pandey}, \citenamefont {Sels},\ and\ \citenamefont
  {Polkovnikov}}]{FloquetEngg2}%
  \BibitemOpen
  \bibfield  {author} {\bibinfo {author} {\bibfnamefont {P.~W.}\ \bibnamefont
  {Claeys}}, \bibinfo {author} {\bibfnamefont {M.}~\bibnamefont {Pandey}},
  \bibinfo {author} {\bibfnamefont {D.}~\bibnamefont {Sels}},\ and\ \bibinfo
  {author} {\bibfnamefont {A.}~\bibnamefont {Polkovnikov}},\ }\bibfield
  {title} {\bibinfo {title} {Floquet-engineering counterdiabatic protocols in
  quantum many-body systems},\ }\href
  {https://doi.org/10.1103/PhysRevLett.123.090602} {\bibfield  {journal}
  {\bibinfo  {journal} {Phys. Rev. Lett.}\ }\textbf {\bibinfo {volume} {123}},\
  \bibinfo {pages} {090602} (\bibinfo {year} {2019})}\BibitemShut {NoStop}%
\end{thebibliography}%
\end{document}